\documentclass[journal,draftcls,onecolumn,12pt,twoside]{IEEEtranTCOM}
\normalsize

\usepackage{cite}

\ifCLASSINFOpdf
   \usepackage[pdftex]{graphicx}
\else
   \usepackage[dvips]{graphicx}
\fi

\usepackage{amssymb,amsfonts}
\usepackage{textcomp}
\usepackage[cmex10]{amsmath}
\usepackage{algorithm}
\usepackage{algpseudocode}
\usepackage[tight,footnotesize]{subfigure}

\usepackage{fixltx2e}
\renewcommand{\baselinestretch}{1.5}
\newtheorem{lemma}{Lemma}
\begin{document}
%
\title{Constrained Partial Group Decoding with Max-Min Fairness for Multi-color Multi-user Visible Light Communication}

%
%
%

\author{Guangtao Zheng, Chen Gong and Zhengyuan Xu
\thanks{This work was supported by Key Program of National Natural Science Foundation of China (Grant No. 61631018) and Key Research Program of Frontier Sciences of CAS (Grant No. QYZDY-SSW-JSC003).}
\thanks{Guangtao Zheng, Chen Gong, and Zhengyuan Xu are with Key Laboratory of
	Wireless-Optical Communications, Chinese Academy of Sciences, School of
	Information Science and Technology, University of Science and Technology of
	China, Hefei, China. Email: zhguangt@mail.ustc.edu.cn,
	\{cgong821, xuzy\}@ustc.edu.cn.}
}

\maketitle

\begin{abstract}
A visible light communication (VLC) system can adopt multi-color light emitting diode (LED) arrays to support multiple users. In this paper, a multi-layer coding and constrained partial group decoding (CPGD) method is proposed to tackle strong color interference and increase the system throughput. After channel model formulation, user information rates are allocated and decoding order for all the received data layers is obtained by solving a max-min fairness problem using a greedy algorithm. An achievable rate is derived under the truncated Gaussian input distribution. To reduce the decoding complexity, a map on the decoding order and rate allocation is constructed for all positions of interest on the receiver plane and its size is reduced by a classification-based algorithm. Meanwhile, the symmetrical geometry of LED arrays is exploited. Finally, the transmitter-user association problem is formulated and solved by a genetic algorithm. It is observed that the system throughput increases as the receivers are slightly misaligned with corresponding LED arrays due to the reduced interference level, but decreases afterwards due to the weakened link gain.
\end{abstract}


%
\IEEEpeerreviewmaketitle

\section{Introduction}
\label{sec:introduction}
\IEEEPARstart{W}{ith} the increasing amount of data transmitted via wireless links, the spectrum shortage has become a critical problem for the next generation communication systems. Visible light communication (VLC) has emerged as a competent supplement~\cite{6011734,6685754} to radio-frequency communication due to its unique advantages, such as large bandwidth and free of electromagnetic radiation. Combining with its other advantages, such as simple transceiver structures due to its intensity modulation-direct detection (IM-DD) method, VLC has attracted extensive research attentions in recent years~\cite{7054450,6844864,5238736,5342315}. 

Light emitting diode (LED) array can be adopted to provide required
illumination and simultaneously increase the system's data rate~\cite{6568880,6825137}. In a multi-user scenario, where each LED and photodiode (PD) pair serves one user, the PD receives not only the desired signal but also the signals from adjacent LEDs as interferences. It is typical that these interferences are strong and thus treating them as noise will penalize the system throughput or even cause a decoding failure. To tackle this problem, interference can be suppressed by designing a resource allocation scheme~\cite{7390880} or aligned~\cite{7418383,6846359} via transmitter-side signal processing. 

One the other hand, it is generally accepted that while a receiver is not interested in decoding the messages from interferers, decoding them is often beneficial for recovering the desired message~\cite{1056307}. Multi-layer coding and group decoder~\cite{gong2011interference} provide an interference cancellation framework that exploits this benefit. At the transmitter, each data stream is split into multiple layers that are encoded individually using independent codebooks, followed by an LED sending the superposition of those layers. At the receiver, under an optimal decoding order, signals are recovered in a successive manner with interference signals partially decoded. 

To control the decoding complexity, the group size is constrained to be a small number, which is called constrained partial group decoder (CPGD)~\cite{gong2011interference,ChenGongCPGD}. CPGD aims to perform  rate optimization under fairness over the achievable region with the optimal decoding order~\cite{7524007}. In a multi-user scenario, the symmetric fairness where all users are constrained with the same rate can guarantee the global rate fairness~\cite{5319748}, while the max-min fairness will improve the sum rate with the fairness maintained in a group. 

A drawback on the CPGD is that the transmitters need feedbacks from the receivers to determine the rate of each layer, which incurs significant communication overhead, especially for a large number of transmitters and users. However, due to the static channel characteristics for VLC~\cite{Chvojka:15}, we can pre-solve the CPGD-related problem of optimal decoding orders and rate allocations at each position, which significantly reduces the online computation complexity. Based on the decoding order, the concept of decoding map is proposed.

The major contributions of this paper are as follows:
\begin{itemize}
	\item  For the multi-color multi-user VLC, we adopt the multi-layer coding and group decoder to perform successive interference cancellation. We obtain an achievable rate of a VLC-multiple access channel (VLC-MAC). 
	\item Due to the static nature of VLC channel, we build a decoding map that stores the decoding order and the corresponding maximal allowable rates for the transmitters at receiver position under the max-min fairness. 
	A classification-based algorithm is proposed to reduce the map size. By exploiting the symmetrical geometry of transmitters, a higher compression ratio can be achieved even at a high transmitter density.
	\item Finally, given the decoding map and users' positions, we formulate the transmitter-user association problem.
	It is observed that receivers at asymmetrical positions will achieve higher sum rate.
\end{itemize}

The remainder of this paper is organized as follows. The system descriptions are give in Section~\ref{sec:des}, on the channel model, multi-layer coding and CPGD. Section~\ref{sec:decoding-map-and-size-reduction} provides the detailed procedures on the CPGD, and the decoding map construction with size reduction based on the symmetric properties. The problem of transmitter-user association and rate allocation is
formulated and solved in Section~\ref{sec:channel-assoc-and-alloc}. Numerical results are provided in Section~\ref{sec:numerical-results}. We conclude this paper in Section~\ref{sec:conclusion}.

\section{System Descriptions}
\label{sec:des}

\subsection{Channel Model}
Consider a multi-color multi-user static VLC channel, with cross-talk at the receiver. The channel gain from the $i^{th}$ transmitter to the $j^{th}$ receiver is denoted as $h_{ji}$ for $i\in\{1,\ldots,N_t\}$ and $j\in\{1,\ldots,N_r\}$, which remains constant during the transmission of length-$N$ codeword block and may change to other states independently afterwards. The signal at the $j^{th}$ receiver at time index $t\in\{1,\ldots,N\}$ is given by
\begin{equation}\label{eq:rx-tx-sig}
	y_j[t]=\gamma\sum_{i=1}^{N_t}h_{ji}x_i[t]+z_j[t],
\end{equation}
where $\gamma$ is the photodetector responsivity and is assumed to be 1 without loss of generality,  $h_{ji}$ is Lambertian channel gain computed by~\cite{ghassemlooy2012optical}
\begin{equation}
	h_{ji}=   \frac{A_r(m_1+1)}{2\pi d_{ji}^2}\cos^{m_1}(\phi_{ji}) 
	T_s(\psi_{ji})g(\psi_{ji})\cos(\psi_{ji}),
\end{equation}
when $0\leq\psi_{ji}\leq\psi_c$ and $h_{ji}=0$ otherwise, where $m_1$ is the Lambertian emission order; $A_r$ is the area of PD; $d_{ji}$ is the distance between transmitter $i$ and receiver $j$; $\phi_{ji}$ is the radiance angle at the transmitter $i$; $\psi_{ji}$ is the incidence angle at the receiver $j$; and $\psi_c$ is the field of view (FOV) at the receiver. Moreover, the concentrator gain $g(\psi_{ji})$ is computed as 
\begin{equation}
	g(\psi_{ji})=\left\{ {\begin{array}{*{20}c}
			\frac{n^2}{\sin^2\psi_c},\hfill & 0\leq\psi_{ji}\leq\psi_c;\\
			{0,} \hfill & \psi_{ji}>\psi_c;
	\end{array}} \right.
\end{equation}
where $n$ is the refractive index. The additive noise term $z_j[t]\sim\mathcal{N}(0,\sigma^2_j)$ in Eq.~\eqref{eq:rx-tx-sig} satisfies Gaussian distribution with zero mean and variance $\sigma^2_j$. Moreover, symbol $x_i[t]$ is sent from the $i$th transmitter and is subject to non-negativity, peak and average power constraints, which are formulated as
\begin{equation}
	x_i\geq 0, x_i\leq A_i, \mathbb{E}[x_i]\leq\varepsilon_i, \forall i=1,\ldots, N_t.
\end{equation}
The optical filter gain is denoted as $T_s(\psi_{ji})$, which is   determined by the corresponding transmitted color band and the receiver-side optical filter. 

\begin{figure}[htbp]
	\centering
	\includegraphics[width=0.55\linewidth]{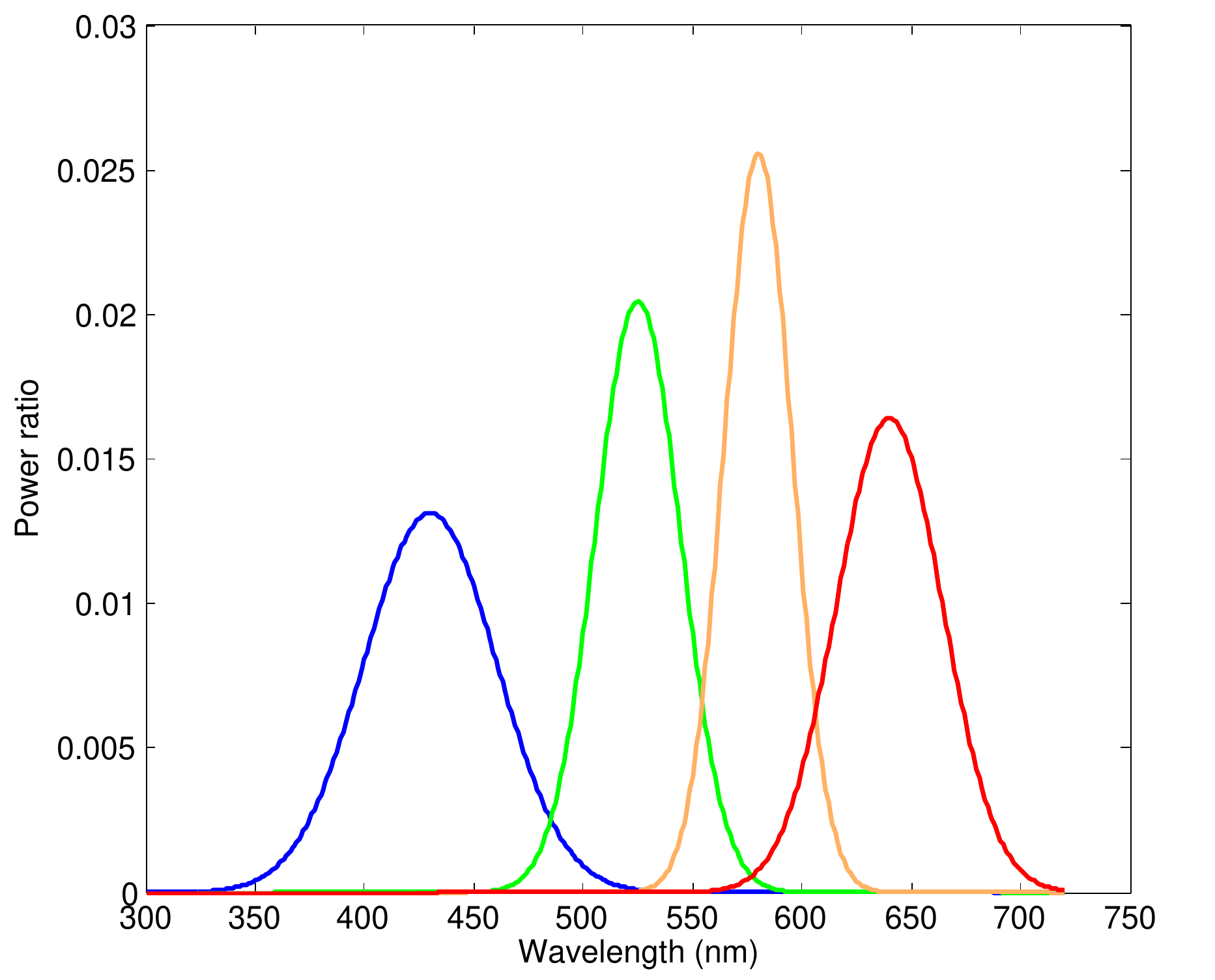}
	\caption{Four color spectra modeled with Gaussian distributions.}
	\label{fig:four-color-spectrum}
\end{figure}

Considering multi-color transmission, we may need to obtain the filtering gain matrix $\mathbf{F}$~\cite{8064179} with its $p$th row and $q$th column element being the power ratio from the transmitter with the $p$th color to the $q$th optical filter.
Letting $N_c$ be the number of colors, we first model the spectrum of the $p$th ($p=1,\ldots,N_c$) color mainly within the wavelength $[\lambda_{p1},\lambda_{p2}]$ with Gaussian distribution~\cite{7728023} $\mathcal{N}(\mu_p,\sigma_p^2)$, where $\mu_p=(\lambda_{p1}+\lambda_{p2})/2$ and $\sigma_p$ is obtained by solving  $\int_{\lambda_{p2}}^{\infty}\mathcal{N}(\mu_p,\sigma_p^2)d\lambda=l_p^r/2$ for leakage probability $l_p^r$ out of band $[\lambda_{p1},\lambda_{p2}]$. Assume that the $q$th optical filter has the passband over wavelength $[\lambda_{q1},\lambda_{q2}]$. Then $\mathbf{F}_{pq}$ can be obtained by integrating the $p$th color spectrum over the band $[\lambda_{q1},\lambda_{q2}]$, i.e., $\mathbf{F}_{pq}=\int_{\lambda_{q1}}^{\lambda_{q2}}\mathcal{N}(\mu_p,\sigma_p^2)d\lambda$.
In the following, we give an example on four colors with wavelength ranges [380nm, 480nm], [500nm, 550nm], [560nm, 600nm], [600nm, 680nm], where each spectrum is modeled by Gaussian distribution with out-band leakage ratios [0.1, 0.2, 0.2, 0.1], respectively. The four color spectra are shown in Figure~\ref{fig:four-color-spectrum}. 
The filtering gain matrix is given in the following
\begin{equation}
	\begin{bmatrix}
		0.900&0.011  &0.000  &0.000 \\ 
		0.011& 0.800 & 0.036 &0.000 \\ 
		0.000&0.027  &  0.800&0.100 \\ 
		0.000&0.000  & 0.050 & 0.900
	\end{bmatrix},
\end{equation}
where each row denotes the power ratio received at four optical filters for each color band. It is observed that the receiver receives not only the signals from the same color band, but also signals from other colors. If the $i$th transmitter uses the $p$th color band and the $j$th receiver uses the $q$th optical filter, then $T_s(\psi_{ji})=\mathbf{F}_{pq}$. 

\subsection{ Input Signals and Layered Encoding}
\label{sec:input-sig-layered-enc}
Multi-layer coding and CPGD are adopted  in the system to decode and cancel the interference successively. At each transmitter, the data stream is split into multiple layers and encoded by independent codebooks with specified rates. The number of layers allocated to transmitter $i$ is denoted as $L_i$, and thus the total number of layers is $L=\sum_{i=1}^{N_t}L_i$. The set of codebooks at transmitter $i$ is denoted as $\mathbf{C}_i\triangleq\{C_{i1},\ldots,C_{iL_i}\}$ for $i=1,\ldots,N_t$. Denoting the signal drawn from codebook $C_{ik}$ at time index $t$ by $x_{ik}[t]$, the received signal in Eq.~\eqref{eq:rx-tx-sig} can be represented by the superposition of multiple layers as follows,
\begin{equation}\label{eq:rx-layer-sig}
	y_j[t]=\sum_{i=1}^{N_t}h_{ji}\sum_{k=1}^{L_i}x_{ik}[t]+z_j[t].
\end{equation} 
The rate $R_i$ at transmitter $i$ can be expressed as $R_i=\sum_{k=1}^{L_i}R_{ik}$, where $R_{ik}$ is the rate for layer $k$ encoded by $C_{ik}$. In the following analysis, we omit the time index $t$ in Eq.~\eqref{eq:rx-layer-sig} for simplicity. Moreover, we use $(i,k)$ to denote the coded layer $k$ of user $i$, for $1\leq i\leq N_t$ and $1\leq k\leq L_i$, and define $\mathcal{K}$ as the ensemble of all layers, i.e., ${\cal K} \triangleq \{(i,k)\}_{1 \leq i \leq N_t, 1 \leq k \leq L_i}$.

Assuming that the power is equally allocated among the layers at each transmitter and the signals of layers are independent of each other,  we have the following constraints:
\begin{equation}\label{eq:ori-constraints}
	x_{ik}\geq 0, x_{ik}\leq A_i/L_i, \mathbb{E}[x_{ik}]\leq\varepsilon_i/L_i, \forall i=1,\ldots, N_t.
\end{equation}
According to~\cite{5238736}, a necessary condition for an optimal input distribution that achieves the capacity is that the average of input signal equals half peak power. Combined with the average power constraint, for each layer, Eq.~\eqref{eq:ori-constraints} can be refined as
\begin{equation}\label{eq:opt-constraints}
	x_{ik}\geq 0, x_{ik}\leq A_i/L_i,\mathbb{E}[x_{ik}]\leq\min\{\frac{\varepsilon_i}{L_i},\frac{A_i}{2L_i}\}, \forall i=1,\ldots, N_t.
\end{equation}

As truncated Gaussian (TG) distribution~\cite{ChaabanVLCbounds}  can be very tight to the capacity at high signal-to-noise-ratio (SNR) that is typical for indoor VLC, we adopt this distribution to approximate the capacity. The TG distribution is determined by a three-tuple parameter $(\mu, \nu, A)$ as follows,
\begin{equation}\label{eq:truncated-gaussian}
	\theta_{\mu,\nu}^A(x)=  \left\{ {\begin{array}{*{20}c}
			{\rho g_{\mu,\nu}(x),} \hfill & x\in[0,A];\\
			{0,} \hfill & \text{otherwise};
	\end{array}} \right.
\end{equation}
where $\rho=(G_{\mu,\nu}(A)-G_{\mu,\nu}(0))^{-1}$, $g_{\mu,\nu}(x)$ is Gaussian distribution with mean $\mu$ and variance $\nu^2$; and $G_{\mu,\nu}(x)$ is the corresponding cumulative Gaussian distribution. For a $(\mu, \nu, A)$-TG distribution, the mean $\hat{\mu}$ and variance $\hat{\nu}^2$ are given by
\begin{equation}\label{eq:tg-mean}
	\hat{\mu}=\nu^2[\theta_{\mu,\nu}^A(0)-\theta_{\mu,\nu}^A(A)]+\mu,
\end{equation}
\begin{equation}
	\hat{\nu}^2=\nu^2[1-A\theta_{\mu,\nu}^A(A)-\hat{\mu}(\theta_{\mu,\nu}^A(0)-\theta_{\mu,\nu}^A(A))].
\end{equation}
For each layer, a $(\mu_{ik},\nu_{ik},A_{ik})$-TG distribution is adopted for the input signal $x_{ik}$, where the TG parameters are delicately selected such that $\hat{\mu}_{ik}\leq\min\{\varepsilon_i/L_i,A_i/2L_i\}$. 

\subsection{Constrained Partial Group Decoder (CPGD)}
We allow each user to decode a subset of the interferers along with the desired messages. To achieve this, each receiver performs a multi-stage group successive decoding by partitioning users' signals into several groups $\mathcal{Q}^j=\{\mathbf{Q}_1^j,\ldots,\mathbf{Q}_{p_{j}}^{j}\}$, and decodes one group via maximum likelihood (ML) criterion, while treating the remaining layers as additive white Gaussian noise (AWGN) in each stage, until the desired signals are decoded. To control the decoding complexity, the size of each group should follow $|\mathbf{Q}_\ell^j|\leq\tau$ for $\forall \ell=1,\ldots,p_j$, hence the name constrained partial group decoder.
More specifically, for any $b$, $1 \leq b \leq L_i$, there exists $m_j\leq p_j$ such that, $(i,b)\in\mathbf{Q}_{m_j}^j$ and $(i,b)\notin\cup_{k=m_j+1}^{p^j}\mathbf{Q}_k^j$, i.e., all signal layers have been decoded before stage $m_j+1$. The CPGD decodes the desired messages from the received signal $y_j$ as follows~\cite{gong2011interference}
\begin{enumerate}
	\item Initialize $m=1$.
	\item Compute  $\Sigma_m^j=\sigma_j^2+\sum_{n=m+1}^{p_j}\sum_{(i,k)\in \mathbf{Q}^j_n}h_{ji}^2\hat{\nu}^2_{ik}$ 
	as the AWGN variance and jointly decode the layers in $\mathbf{Q}_m^j$ via ML decoding.
	\item Update $y_j\leftarrow y_j-\sum_{(i,k)\in\mathbf{Q}_m^j}h_{ji}\hat{x}_{ik}$, where $\hat{x}_{ik}$ is decoded symbol of $x_{ik}$.
	\item If $m=m_j$, stop and output the information of desired layers. Otherwise, update $m \leftarrow m+1$ and  go to step 2.
\end{enumerate}

The optimal group partitioning for each user depends on the objective function to be optimized. In this paper, the max-min objective function among the rates of all users is adopted. Usually this optimization problem is solved in a distributed manner, where each receiver obtains  the optimal decoding order and the locally maximal rates achievable for the transmission layers. Then, the transmitters need feedbacks from the receivers to obtain the global rate allocation via taking the minimum of all locally obtained rates for each transmission layer. A decoding outage may otherwise be declared if some layer's rate exceeds its corresponding channel capacity.
More details on CPGD can be found in~\cite{gong2011interference} and references therein.

\section{Decoding Map and Size Reduction}
\label{sec:decoding-map-and-size-reduction}
Assuming accurate knowledge on the receiver's position, we can pre-compute the decoding order and the corresponding rate allocation at each position of interest. Such results can be obtained by solving the corresponding rate maximization problem. In this section, we first calculate the achievable rate of a VLC multiple access channel, which lays the foundation for solving the formulated rate allocation problem. Then we introduce the concept of decoding map and provide an algorithm to obtain it. To effectively represent the decoding map and show the symmetric properties, a size reduction algorithm is proposed.
\subsection{Achievable Rate Calculation}
\label{sec:achievable-rate-calc}
Denote $\mathcal{V}$ and $\mathcal{G}$ as two subsets of $\mathcal{K}$ with ${\cal V} \bigcap {\cal G} = \emptyset$, where the former represents the signal index set and the latter is treated as interference. As will be shown in Section~\ref{sec:rate-max-problem}, rate  $R(\mathcal{V},\mathcal{G})$ is fundamental in solving  problem~\eqref{eq:rate-max-problem-main}. 
Due to the signal's non-negativity with peak and average power constraints, the closed form of a VLC channel capacity is still open ~\cite{ChaabanVLCbounds}. However, a good closed-form achievable rate approximation that approaches the capacity can be obtained using the truncated Gaussian distribution given in Section~\ref{sec:input-sig-layered-enc}. 

Since the size of set $\mathcal{V}$ is at least one, we aim to solve the achievable rate of a VLC multiple access channel. 
Define the rate for receiver $j$ as 
$R_j(\mathcal{V},\mathcal{G})=I(X_{\mathcal{V}};Y_j)$, where $Y_j$ is defined as
\begin{equation}
	Y_j=\sum_{(i,k)\in \mathcal{V}}h_{ji}x_{ik}+\sum_{(i,k)\in \mathcal{G}}h_{ji}x_{ik}+z_j.
\end{equation}
To calculate $R_j(\mathcal{V},\mathcal{G})$, we begin with
\begin{equation}
	\begin{aligned}
		R_j(\mathcal{V},\mathcal{G})&=h(X_{\mathcal{V}})-h(X_{\mathcal{V}}|Y_j)\\
		&=\sum_{(i,k)\in\mathcal{V}}h(X_{ik})-h(X_{\mathcal{V}}|Y_j).
	\end{aligned}
\end{equation}
Moreover, letting $X_{{\cal V},m}$ denote the $m$th elements in set ${\cal V}$, we expand $h(X_{\mathcal{V}}|Y_j)$ with chain rule~\cite{cover2012elements} to get
\begin{equation}
	\begin{aligned}
		h(X_{\mathcal{V}}|Y_j)
		&=\sum_{m=1}^{p}h(X_{\mathcal{V},i}|Y_j,X_{\mathcal{V},1},\ldots,X_{\mathcal{V},m-1}),
	\end{aligned}
\end{equation}
where $p$ is the number of layers decoded.
We then invoke the following lemma~\cite{thomas1987feedback}:
\begin{lemma}
	Let $X_1, X_2,\ldots, X_k$ be an arbitrary set of random variables with mean $\mathbf{\mu}$ and covariance matrix $\mathbf{K}$. Let $\mathcal{S}$ be any subset of $\{1,2,\ldots,k\}$ and $\bar{\mathcal{S}}$ be its complement. Then 
	\begin{equation}
		h(X_{\mathcal{S}}|X_{\bar{\mathcal{S}}})\leq h(X_{\mathcal{S}}^{*}|X_{\bar{\mathcal{S}}}^{*}),
	\end{equation}
	where $(X_1^*, X_2^*,\ldots, X_k^*)\sim \mathcal{N}(\mathbf{\mu},\mathbf{K})$.$\hfill \Box$
\end{lemma}
According to Lemma~$1$, $h(X_{\mathcal{V}}|Y_j)$ can be upper bounded by $h(X_{\mathcal{V}}^*|Y_j^*)$, where $(X_{\mathcal{V}}^*,Y_j^*)$ is jointly Gaussian with the same covariance matrix as that of $(X_{\mathcal{V}},Y_j)$. 
Then, we have
\begin{equation}\label{eq:condition-entropy-upper-bound}
	h(X_{\mathcal{V}}|Y_j)\leq\sum_{m=1}^{p}h(X^*_{\mathcal{V},m}|Y^*_j,X^*_{\mathcal{V},1},\ldots,X^*_{\mathcal{V},m-1}),
\end{equation}
where $(X^*_{\mathcal{V},m},X^*_{\mathcal{V},1},\ldots,X^*_{\mathcal{V},m-1},Y^*_j)$ are jointly Gaussian with covariance matrix $\mathbf{K}_m$.

We calculate covariance matrix $\mathbf{K}_m$ under the group size constraint $|X_{\mathcal{V}}|\leq\tau$, according to CPGD. We write $X_{\mathcal{V}}$ in a sequence $X_{\mathcal{V},1},\ldots,X_{\mathcal{V},p}$, where $(\mathcal{V},m) , \forall m=1,\ldots,p$ denotes the user layer index in $\mathcal{V}$, with the corresponding variance  $\hat{\nu}^2_{\mathcal{V},1},\ldots,\hat{\nu}^2_{\mathcal{V},p}$ for $p\leq\tau$. Similarly, we write $X_{\mathcal{G}}$ in a sequence $X_{\mathcal{G},1},\ldots,X_{\mathcal{G},q}$, where $(\mathcal{G},m) , \forall m=1,\ldots,q$ is an element in $\mathcal{G}$, with the corresponding variance $\hat{\nu}^2_{\mathcal{G},1},\ldots,\hat{\nu}^2_{\mathcal{G},p}$. Hence $\mathbf{K}_m$ can be  written in Eq.~\eqref{eq:covariance-matrix}.
\begin{figure*}
	\begin{equation}\label{eq:covariance-matrix}
		\mathbf{K}_m=	\begin{bmatrix}
			\hat{\nu}^2_{\mathcal{V},m} & 0 & \ldots &  0& h_{j(\mathcal{V},i)}\hat{\nu}^2_{\mathcal{V},m}\\ 
			0& \hat{\nu}^2_{\mathcal{V},1} & \dots & 0 &h_{j(\mathcal{V},1)}\hat{\nu}^2_{\mathcal{V},1} \\ 
			\vdots  & \vdots & \ddots  & \vdots & \vdots\\ 
			0& 0 & \ldots & \hat{\nu}^2_{\mathcal{V},m-1} & h_{j(\mathcal{V},m-1)}\hat{\nu}^2_{\mathcal{V},m-1}\\ 
			h_{j(\mathcal{V},m)}\hat{\nu}^2_{\mathcal{V},m}&h_{j(\mathcal{V},1)}\hat{\nu}^2_{\mathcal{V},1}  & \ldots &h_{j(\mathcal{V},m-1)} \hat{\nu}^2_{\mathcal{V},m-1} & \sum_{\ell=1}^ph^2_{j(\mathcal{V},\ell)}\hat{\nu}^2_{\mathcal{V},\ell}+\sum_{\ell=1}^qh^2_{j(\mathcal{G},\ell)}\hat{\nu}^2_{\mathcal{G},\ell}+\sigma^2_j
		\end{bmatrix}.
	\end{equation}
\end{figure*}
According to Eq.(40) in~\cite{thomas1987feedback}, $h(X^*_{\mathcal{V},m}|Y^*_j,X^*_{\mathcal{V},1},\ldots,X^*_{\mathcal{V},m-1})$ can be obtained by
\begin{equation}
	h(X^*_{\mathcal{V},m}|Y^*_j,X^*_{\mathcal{V},1},\ldots,X^*_{\mathcal{V},m-1})=\frac{1}{2}\log 2\pi e(\hat{\nu}^2_{\mathcal{V},m}-\mathbf{b}\mathbf{S}^{-1}\mathbf{b}^T),
\end{equation}
where $\mathbf{b}=[0,\ldots,0,h_{j(\mathcal{V},i)}\hat{\nu}^2_{\mathcal{V},m}]$; $\mathbf{S}$ is the matrix $\mathbf{K}_m$ with its first row and column being deleted. By exploiting the sparsity of the matrix $\mathbf{K}_m$, we can explicitly express each conditional entropy at the right hand side of Eq.~\eqref{eq:condition-entropy-upper-bound} as
\begin{equation}
		h(X^*_{\mathcal{V},m}|Y^*_j,X^*_{\mathcal{V},1},\ldots,X^*_{\mathcal{V},m-1})=\frac{1}{2}\log 2\pi e(\hat{\nu}^2_{\mathcal{V},m}
		-\frac{h^2_{j(\mathcal{V},m)}\hat{\nu}^4_{\mathcal{V},m}}{\sum_{\ell=m}^ph^2_{j(\mathcal{V},\ell)}\hat{\nu}^2_{\mathcal{V},\ell}+\sum_{\ell=1}^qh^2_{j(\mathcal{G},\ell)}\hat{\nu}^2_{\mathcal{G},\ell}+\sigma^2_j}),
\end{equation}
and thus $h(X_{\mathcal{V}}|Y_j)$ can be upper bounded by
\begin{equation}
			h(X_{\mathcal{V}}|Y_j)\leq \frac{p}{2}\log 2\pi e+
		\sum_{m=1}^p\frac{1}{2}\log\Big(\hat{\nu}^2_{\mathcal{V},m}- \frac{h_{j(\mathcal{V},m)}^2\hat{\nu}^4_{\mathcal{V},m}}{\sum_{\ell=m}^ph^2_{j(\mathcal{V},\ell)}\hat{\nu}^2_{\mathcal{V},\ell}+\sum_{\ell=1}^qh^2_{j(\mathcal{G},\ell)}\hat{\nu}^2_{\mathcal{G},\ell}+\sigma^2_j}\Big).	
\end{equation}
Therefore, an achievable rate at receiver $j$  can be computed as follows,
\begin{equation}
	\label{eq:achievable-rate-vlc-mac}
	R_j(\mathcal{V},\mathcal{G})=\frac{1}{2}\sum_{m=1}^p\Big[\log\nu_{\mathcal{V},m}^2-\log\Big(\hat{\nu}^2_{\mathcal{V},m}-
 \frac{h_{j(\mathcal{V},m)}^2\hat{\nu}^4_{\mathcal{V},m}}{\sum_{\ell=m}^ph^2_{j(\mathcal{V},\ell)}\hat{\nu}^2_{\mathcal{V},\ell}+\sum_{\ell=1}^qh^2_{j(\mathcal{G},\ell)}\hat{\nu}^2_{\mathcal{G},\ell}+\sigma^2_j}\Big) \Big]-\sum_{m=1}^p\phi_{\mathcal{V},m},	
\end{equation}
where $\phi_{ik}=\log(\rho_{ik})+\frac{1}{2}\big((A_{ik}-\mu_{ik})\theta_{\mu_{ik},\nu_{ik}}^{A_{ik}}(A_{ik})+\mu_{ik}\theta_{\mu_{ik},\nu_{ik}}^{A_{ik}}(0)\big)$.

\subsection{Problem Statement}
\label{sec:rate-max-problem}
Assume that at position $\mathbf{x}=[x_1,x_2,x_3]$, the group decoding order is denoted as  $\mathcal{Q}^{\mathbf{x}}=\{\mathbf{Q}_1^{\mathbf{x}},\ldots,\mathbf{Q}_{p_\mathbf{x}}^{\mathbf{x}}\}$. The optimal decoding order $\hat{\mathcal{Q}}^{\mathbf{x}}$ is obtained via solving the following optimization problem,
\begin{equation}
	\label{eq:rate-max-problem-main}
	\hat{\mathcal{Q}}^{\mathbf{x}}=\arg\underset{\mathcal{Q}^{\mathbf{x}}\in\mathcal{\underline{Q}}}{\max }\psi(\mathcal{Q}^{\mathbf{x}}),
\end{equation} 
where $\mathcal{\underline{Q}}$ is the set of all legitimate decoding order and $\psi(\mathcal{Q}^{\mathbf{x}})$ is the max-minimum rate of all layers given by
\begin{equation}\label{eq:rate-max-problem-sub}
	\psi(\mathcal{Q}^{\mathbf{x}})=\left\{ {\begin{array}{*{20}c}
			\text{max}\hfill & \underset{(i,k)\in\mathcal{K}^{\mathbf{x}}}{\min}r_{(i,k)}^{\mathbf{x}}\\
			{\text{s.t.}} \hfill & \mathbf{r}_{\mathbf{Q}_m^{\mathbf{x}}}\in \mathcal{C}(\mathbf{Q}_m^{\mathbf{x}},\cup_{j=m+1}^{p_\mathbf{x}}\mathbf{Q}_j^{\mathbf{x}}) \\
			& \forall m\in \{1,\ldots,p_{\mathbf{x}}\}
	\end{array}} \right.,
\end{equation}
where rate vector  $\mathbf{r}_{\mathbf{Q}_m^{\mathbf{x}}}\triangleq[r_{(i,k)}^{\mathbf{x}}]$,$\forall(i,k)\in\mathbf{Q}_m^{\mathbf{x}}$;  $r_{(i,k)}^{\mathbf{x}}$ represents the maximum allowable rate of the layer $(i,k)$ obtained at the position $\mathbf{x}$;  $\mathcal{C}(\mathbf{Q}_m^{\mathbf{x}},\cup_{j=m+1}^{p_\mathbf{x}}\mathbf{Q}_j^{\mathbf{x}})$ represents the achievable rate region of a VLC-MAC treating $\cup_{j=m+1}^{p_\mathbf{x}}\mathbf{Q}_j^{\mathbf{x}}$ as noise; $p_\mathbf{x}$ is the number of layer groups decoded
at position $\mathbf{x}$. By solving problems~\eqref{eq:rate-max-problem-main} and~\eqref{eq:rate-max-problem-sub}, we can obtain the maximum achievable rate for each layer and the corresponding optimal decoding order.

For this purpose, we first invoke the rate increment margin defined as follows~\cite{7524007},
\begin{equation}
	\label{eq:rate-max-sub-solution}
	\Delta_{\mathbf{x}}(\mathcal{V},\mathcal{G},\mathbf{R})\triangleq\underset{\mathcal{D}\neq\emptyset,\mathcal{D}\subset\mathcal{V}}\min\frac{R_{\mathbf{x}}(\mathcal{D},\mathcal{G})-\|\mathbf{R}_{\mathcal{D}}\|_1}{|\mathcal{D}|},
\end{equation}
where $\mathbf{R}_{\mathcal{D}}$ contains the rates selected by $\mathcal{D}$ from any decodable rate vector $\mathbf{R}$; $\mathcal{V}$ acts as the signal layer set and  $\mathcal{G}$ is treated as the noise layer set with  $\mathcal{V}$, $\mathcal{G}\subset\mathcal{K}$ and 
$\mathcal{V}\cap\mathcal{G}=\emptyset$; $|\cdot|$ denotes the size of a set and $\|\cdot\|_1$ denotes the norm-1 operation. The definition means at position $\mathbf{x}$, if the rate of layers in $\mathcal{V}$ increases beyond $\Delta_{\mathbf{x}}(\mathcal{V},\mathcal{G},\mathbf{R})$, then decoding $\mathcal{V}$ while treating $\mathcal{G}$ as noise may cause an outage. Thus given the decoding order $\{\mathbf{Q}_m^{\mathbf{x}}\}_{m=1}^{p_{\mathbf{x}}}$, for $1\leq m\leq p_{\mathbf{x}}$, the maximum achievable rate allocated to $\mathbf{Q}^{\mathbf{x}}_m$ is $\Delta_{\mathbf{x}}(\mathbf{Q}^{\mathbf{x}}_m,\cup_{\ell>m}\mathbf{Q}^{\mathbf{x}}_\ell,\mathbf{0})$.


Problem ~\eqref{eq:rate-max-problem-main} can be optimally solved by a greedy method under Gaussian signaling, where the achievable rate function  $R_{\mathbf{x}}(\mathcal{V},\mathcal{G})$ exhibits submodularity and transitivity which play the key role in guaranteeing the optimality, shown as follows~\cite{6280254},
\begin{align}
	&R_{\mathbf{x}}(\mathcal{D}_1,\mathcal{G})+R_{\mathbf{x}}(\mathcal{D}_2,\mathcal{G})\geq  R_{\mathbf{x}}(\mathcal{D}_1\cap\mathcal{D}_2,\mathcal{G})+R_{\mathbf{x}}(\mathcal{D}_1\cup\mathcal{D}_2,\mathcal{G}),
	\\
	&R_{\mathbf{x}}(\mathcal{U},\mathcal{G})+R_{\mathbf{x}}(\mathcal{V},\mathcal{G}\cup\mathcal{U})=R_{\mathbf{x}}(\mathcal{U}\cup\mathcal{V},\mathcal{G}),
\end{align}
for disjoint sets $\mathcal{U}$, $\mathcal{V}$, $\mathcal{G}$.
Although due to the non-Gaussian input distributions adopted in this paper, the optimality of original decoding order may not be guaranteed, we still adopt it as a solution with low computational complexity. We will perform the successive decoding and rate allocation as those in the Gaussian case.

\subsection{Decoding Map}
\label{sec:decoding-map}
As the decoding order of group decoding is discrete, we can quantize the user location in the feasible space into various regions, and presume that the decoding order in the center of each region may well represent that for users in that region if the region size is sufficiently small. Based on this, we introduce the concept of decoding map, solve problem~\eqref{eq:rate-max-problem-main} at each position of the map and store the decoding order with the allocated rates for the multi-color multi-user VLC system under consideration,  assuming layered encoding and CPGD for interference cancellation. Once knowing the user position, we can search the map to obtain the decoding orders and rate allocations to configure the transmitter and receiver, such that the online computation overhead can be significantly reduced.

Besides the decoding order of the user layer, we may also need to figure out the users that each transmitter serves, i.e. the transmitter-user association. To compute the transmitter-user association for each user position, we may need to calculate the achievable rate for each signal layer at all possible receivers. To achieve this, we conduct the max-min rate allocation procedure until all signal layers are decoded, as shown in Algorithm~\ref{alg:dec-order-pos-algo}.
\begin{algorithm}[htbp]
	\caption{Compute the decoding order at $\mathbf{x}=(x_1,x_2,x_3)$}
	\label{alg:dec-order-pos-algo}
	\begin{algorithmic}[1]
		\State {Given $\mathcal{D}=\mathcal{K}_{\mathbf{x}}$, $\mathcal{G}=\emptyset$, $\ell=1$, $p_\mathbf{x}=0$ and channel gain $\mathbf{h}_\mathbf{x}$;}
		\Repeat
		\State { $\delta_\ell=\underset{\mathcal{V}\neq\emptyset,\mathcal{V}\subset\mathcal{D},|\mathcal{V}|\leq\tau}{\min}R_\mathbf{x}(\mathcal{V},\mathcal{G})/|\mathcal{V}|$;}
		\State{$\mathcal{G}_\ell^{\mathbf{x}}=\underset{\mathcal{V}\neq\emptyset,\mathcal{V}\subset\mathcal{D},|\mathcal{V}|\leq\tau}{\arg\min}R_\mathbf{x}(\mathcal{V},\mathcal{G})/|\mathcal{V}|$;}
		\State{$\mathcal{D}\leftarrow\mathcal{D}\backslash\mathcal{G}_{\ell}^\mathbf{x}$ and $\mathcal{G}\leftarrow\mathcal{G}\cup\mathcal{G}_{\ell}^\mathbf{x}$;}
		\State{$r_{(i,k)}^{\mathbf{x}}=\delta_\ell,\ \forall (i,k)\in\mathcal{G}_\ell^{\mathbf{x}},\ p_\mathbf{x}=p_\mathbf{x}+1$;}
		\State{$\ell=\ell+1$;}
		\Until {$D=\emptyset$}
		\State{Set $\mathbf{Q}_{m}^{\mathbf{x}}\leftarrow\mathcal{G}_{\ell-m}^{\mathbf{x}},1\leq m\leq p_{\mathbf{x}} 
			$;}
		\State{Output $\mathcal{Q}^{\mathbf{x}}=\{\mathbf{Q}_1^{\mathbf{x}},\ldots,\mathbf{Q}_{p_{\mathbf{x}}}^{\mathbf{x}}\},\mathbf{r}^{\mathbf{x}}=\{r_{(i,k)}^{\mathbf{x}}|\forall(i,k)\in\mathcal{Q}^{\mathbf{x}}\}$.}
	\end{algorithmic}
\end{algorithm}

In Algorithm~\ref{alg:dec-order-pos-algo}, $\mathcal{K}_{\mathbf{x}}=\{(i,k)|\mathbf{h}_{\mathbf{x}}[i]\neq 0, i=1,\ldots,N_t\}$ denotes the set of all layers detectable at position $\mathbf{x}$ , since some layers may not be detected because of the constraint of FOV or being completely filtered out by a mismatched optical filter. The algorithm runs in a greedy manner such that in each step, it finds the layer set that reaches the boundary of the achievable region of a VLC-MAC.

For each color, we sample the positions uniformly among the possible positions of users. At each position, we run Algorithm~\ref{alg:dec-order-pos-algo} and store the achievable rates with the decoding order of all received layers. The online computation merely involves the table look-up operation, which significantly reduces the computational complexity especially in terms of the decoding order determination.
\subsection{Size Reduction for the Decoding Map}
\label{sec:size-reduction-algo}
The size of a decoding map is typically large especially for a large number of transmitters and small distance between the sampling points in the decoding map. This will pose challenges to storage and searching. To reduce the decoding map size, we have to note that finding the rate for each layer under a specified decoding order is much easier than jointly obtaining the decoding order and rate allocation. This means that offline storing the decoding order and online computation of rate allocation can reduce the storage complexity while not introducing a large online computation load. Moreover, the decoding orders are the same for several positions sufficiently close to each other. Even the decoding orders are different, the corresponding rates bear a very small difference such that the decoding order of one position can well represent those at other positions, i.e., the two positions can be classified into the same category.

The size reduction falls into finding a classification algorithm in the space. To achieve this, we first define a normalized distance between two positions $\mathbf{x}_1$ and $\mathbf{x}_2$,
\begin{equation}\label{eq:normalized-distance}
	d(\mathbf{x}_1,\mathbf{x}_2)\triangleq \|\mathbf{r}_{\mathbf{x}_2}-\psi(\mathbf{h}_{\mathbf{x}_2},\sigma_{\mathbf{x}_2}^2,\mathcal{Q}^{\mathbf{x}_1})\|_2/\|\mathbf{r}_{\mathbf{x}_2}\|_2,
\end{equation}
which characterizes the normalized variation between the optimal rate allocation vector at position $\mathbf{x}_2$ and the rate allocation vector $\psi(\mathbf{h}_{\mathbf{x}_2},\sigma_{\mathbf{x}_2}^2,\mathcal{Q}^{\mathbf{x}_1})$ at position $\mathbf{x}_2$ using the decoding order $\mathcal{Q}^{\mathbf{x}_1}$ at $\mathbf{x}_1$. Rate vector $\psi(\mathbf{h}_{\mathbf{x}_2},\sigma_{\mathbf{x}_2}^2,\mathcal{Q}^{\mathbf{x}_1})$ can be obtained from rate allocation according to Eq.~\eqref{eq:rate-max-sub-solution} under decoding order $\mathcal{Q}^{\mathbf{x}_1}=\{\mathbf{Q}_1^{\mathbf{x}_1},\ldots,\mathbf{Q}_{p_{\mathbf{x}_1}}^{\mathbf{x}_1}\}$. The size reduction algorithm is given in Algorithm~\ref{alg:size-reduced-algo}, which reduces the decoding map size of each color.
\begin{algorithm}[htbp]
	\begin{algorithmic}[1]
		\State {Set $K$=1, $\ell=1$, initialize $\tau_{\text{diff}}$, $\tau_{\text{loss}}$, $\mathbf{N}_K=\{\mathcal{N}_k|k=1,\ldots,K\}$;}
		\Repeat
		\For{$k=1:K$}
		\State{$\mathcal{N}_k=\mathbf{N}_K(k)$;}
		\State{$\mathbf{d}_k(i)=\frac{1}{|\mathcal{N}_k|}\sum_{j\in\mathcal{N}_k}d(\mathbf{x}_i,\mathbf{x}_j),\forall i\in\mathcal{N}_k$;}
		\If{$\max(\mathbf{d}_k)>\tau_{\text{loss}}$}
		\Repeat
		\State{$\xi=\mathbf{d}_k(1)$;}
		\State{$\tilde{\mathcal{N}}_\ell=\{i\big| \text{abs}(\xi-\mathbf{d}_k(i))<\tau_{\text{diff}},i\in\mathcal{N}_k\}$;}
		\State{$\mathcal{N}_k=\mathcal{N}_k\backslash\tilde{\mathcal{N}}_\ell$, $\mathbf{d}_k=\mathbf{d}_k\backslash\mathbf{d}_k(\tilde{\mathcal{N}}_\ell)$;}
		\State{$\ell=\ell+1$;}
		\Until{$\mathcal{N}_k=\emptyset$}
		\Else
		\State{$\tilde{\mathcal{N}}_\ell=\mathcal{N}_k$;}
		\State{$\ell=\ell+1$;}
		\EndIf
		\EndFor
		\State{$K=\ell-1$, $\mathbf{N}_K=\{\tilde{\mathcal{N}}_k|k=1,\ldots,K\}$;}
		\Until{$K$ does not change}
		\State{Output $K$ and $\mathbf{N}_K$.}
	\end{algorithmic}
	\caption{Size reduction of a decoding map}
\label{alg:size-reduced-algo}
\end{algorithm}

In Algorithm~\ref{alg:size-reduced-algo}, $\tau_{\text{diff}}$ is the threshold that determines whether two positions are in the same category, while $\tau_{\text{loss}}$ is the maximal tolerable loss and  greater than $\tau_{\text{diff}}$ in general. We label each position sequentially using integers as the indexes. Let $\mathcal{N}_k$ denote the index set for the $k$th category, and $\mathbf{N}_K$ denote the entire index sets.

The algorithm starts with category number $K=1$, which implies that $\mathcal{N}_k$ contains all points. In each category, we compute the average distance of each point to all the other points. If the maximum average distance is greater than the predefined threshold $\tau_{\text{loss}}$, we need to further divide this category into smaller ones. The partition proceeds in such a way that the points with the difference smaller than the given threshold  $\tau_{\text{diff}}$ are put into a new category. On the other hand, if the maximal average distance is lower than $\tau_{\text{loss}}$, no further partition is needed and the algorithm proceeds to evaluate the next category. The algorithm stops when in all categories no further partition is needed or the category number $K$ does not change.

The algorithm will terminate in finite steps and is effective in reducing the size of a decoding map and producing clear boundaries for all categories. However, as the number of LEDs increases in a fixed area, it becomes more difficult to reduce the size as adjacent transmitter positions bear more variations. 
\subsection{Symmetry Constraint and Size Reduction}
\label{sec:symmetry-constraint-size-reduction}
As we observe from Algorithm~\ref{alg:dec-order-pos-algo} and Eq.~\eqref{eq:achievable-rate-vlc-mac}, the decoding order is determined by the link gain $\mathbf{h}_{\mathbf{x}}$ if the noise variances at all positions are the same. Moreover, if the link gain vector ${\bf h}_{\bf x}$ is the permutation of another, then the decoding order can be obtained via certain permutation and we don't need to run Algorithm~\ref{alg:dec-order-pos-algo}  again. Thus, the computational complexity to obtain the decoding map can be significantly reduced.

If the transmitters are uniformly spaced and arranged in a regular shape, e.g., a square which is typical for indoor lighting, then the channel gains exhibit certain symmetry properties with respect to the transmitters. By exploiting these symmetry properties, we can compute the decoding order directly at other rotated or reflected positions based on the results of the current position. 

More specifically, given the square geometry and $N_h\times N_v$ index matrix $\mathbf{A}$, we use $\mathbf{A}(i,j)=\pi_{\mathbf{A}}(i,j)$ to denote the transmitter index at the $i$th row and $j$th column of the LED array, where $\pi_{\mathbf{A}}(i,j)$ is the function that maps position $(i,j)$ into the index. Conversely, $(i,j)=\pi_{\mathbf{A}}^{-1}(\mathbf{A}(i,j))$ computes the position  given the transmitter index.  If we know the decoding layer $(i,k)$ at a certain stage, the corresponding layer at the symmetrical position is $(\mathbf{A}_p(\pi_{\mathbf{A}}^{-1}(i)),k)$, where $\mathbf{A}_p$ is the matrix generated by $\mathbf{A}$ in the following three ways: (1)
$\mathbf{A}_p(i,j)=\mathbf{A}(N_h+1-j,N_v+1-i)$ ($N_h=N_v$ for a square geometry), if two points are symmetrical with respect to (w.r.t) the diagonal; (2) $\mathbf{A}_p(N_h-i+1,j)=\mathbf{A}(i,j)$ if two points are symmetrical w.r.t the horizontal center line; (3) $\mathbf{A}_p(i,N_h-j+1)=\mathbf{A}(i,j)$ if two points are symmetrical w.r.t. the vertical center line.

\begin{figure}[htbp]
	\centering
	\includegraphics[width=0.35\linewidth]{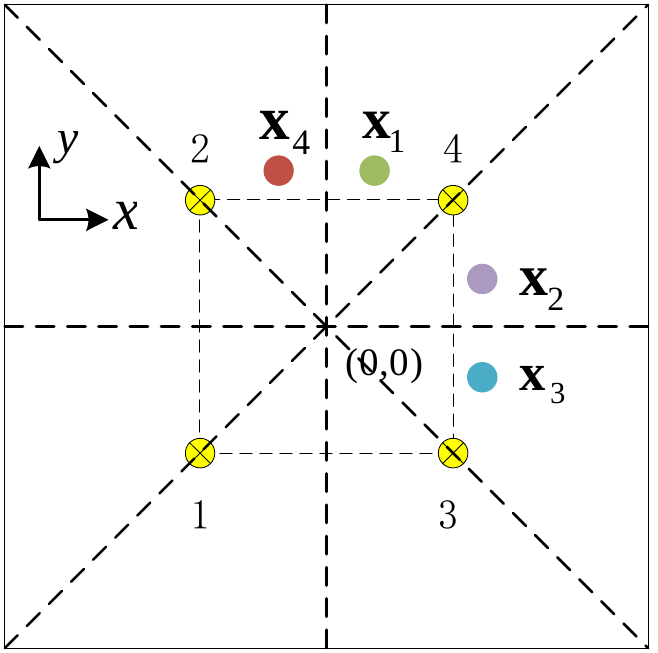}
	\caption{Illustration of exploiting the symmetrical geometry of transmitters to obtain the decoding order at one place based on that at symmetrical positions.}
	\label{fig:decodingsymm}
\end{figure}

Figure~\ref{fig:decodingsymm} gives an example. For simplicity, assuming only one LED with single layer at each position, then the index matrix is given by $\mathbf{A}=[2,4;1,3]$. At point $\mathbf{x}_1$, we have  $\mathcal{Q}^{\mathbf{x}_1}=\{(4,1),(2,1),(3,1),(1,1)\}$. Then at point $\mathbf{x}_2$, which is symmetrical to $\mathbf{x}_1$ w.r.t. $y=x$, we first obtain $\mathbf{A}_p=[3,4;1,2]$, apply the transformation $(\mathbf{A}_p(\pi_{\mathbf{A}}^{-1}(i)),k)$, and get the decoding order $\mathcal{Q}^{\mathbf{x}_2}=\{(4,1),(3,1),(2,1),(1,1)\}$. 
Similarly, at point $\mathbf{x}_4$ which is symmetrical w.r.t. $y=0$, we obtain $\mathcal{Q}^{\mathbf{x}_4}=\{(2,1),(4,1),(1,1),(3,1)\}$ with $\mathbf{A}_p=[4,2;3,1]$. Given $\mathcal{Q}^{\mathbf{x}_2}$, we obtain at point $\mathbf{x}_3$, which is symmetrical w.r.t. $x=0$, the decoding order $\mathcal{Q}^{\mathbf{x}_3}=\{(3,1),(4,1),(1,1),(2,1)\}$ with $\mathbf{A}_p=[1,3;2,4]$. By doing the above computation iteratively, we only need to compute $\frac{1}{8}$ of the original space to get the entire decoding map.

\section{Transmitter-User Association with Rate Allocation}
\label{sec:channel-assoc-and-alloc}
Note that in the previous Section, different user associations may give different global rate allocation schemes, some of which will have higher sum  rates than others. To further enhance the transmission rate, the LED-PD transceiver pair association should be optimized. 

We aim to maximize the sum rate of the system. Given the decoding order $\mathcal{Q}^j$ of the $j$th ($j=1,\ldots,N_r$) user, which can be obtained from the decoding map, and the corresponding local rate allocation $\mathbf{r}^j$, we can formulate the optimal transmitter-user association and rate allocation as follows,
\begin{align}
	{\max} & \sum\limits_{i=1}^{N_t}\sum\limits_{j=1}^{N_r}\sum\limits_{\ell=1}^{L_i}\bar{\mathbf{r}}\Big[\eta_{ji}(\sum_{b=1}^{i-1}L_b+\ell)\Big],\label{eq:chan-assoc-rate-alloc-problem}\\
	{\text{s.t.}} \hfill & \sum\limits_{j=1}^{N_r}\eta_{ji}=1,\forall 1 \leq i \leq N_t;\label{eq:unique-rx-constraint}\\	
	& \sum\limits_{i=1}^{N_t}\eta_{ji}=1,\forall 1 \leq j \leq N_t;\label{eq:unique-tx-constraint}\\
	&m_j=\varphi(\boldsymbol{\eta}_j,\mathcal{Q}^{j}),\forall j=1,\ldots,N_r;\label{eq:compute-subset-index}\\
	&\mathbf{r}^j[\Xi(\cup_{k=m_j+1}^{p^j}\mathbf{Q}_k^j)]=+\infty,\forall j=1,\ldots,N_r;\label{eq:discard-layer-infty}\\
	&\bar{\mathbf{r}}[k]=\underset{j}{\min\ } \mathbf{r}^j[k], k=1,\ldots,\sum_{\ell=1}^{N_t}L_\ell;\\
	&\bar{\mathbf{r}}[0]=0\label{eq:initial-constraint}.
\end{align}
In the above problem formulation,  Eq.~\eqref{eq:unique-rx-constraint} means that a transmitter serves only one user, where $\eta_{ji}\in\{0,1\}$;  Eq.~\eqref{eq:unique-tx-constraint} means a user receives the signal from only one transmitter. The function $m_j=\varphi(\boldsymbol{\eta}_j,\mathcal{Q}^{j})$ in Eq.~\eqref{eq:compute-subset-index} computes the subset's index given $\boldsymbol{\eta}_j$ which is the $j$th row of $\boldsymbol{\eta}$, such that $\mathbf{Q}_{m_j}^j$ contains at least one signal layer and the subsets after $m_j$ contain no signal layers. Formally, it is equivalent to the following,
\begin{align}
	&i=\underset{k}{\arg}\{\eta_{jk}=1\}; \\
	&\exists b=1,\ldots,L_i, (i,b)\in\mathbf{Q}_{m_j}^j;\\
	&\forall b=1,\ldots,L_i, (i,b)\notin\cup_{k=m_j+1}^{p^j}\mathbf{Q}_k^j.
\end{align}
Furthermore, Eq.~\eqref{eq:discard-layer-infty} sets the rates of discarded layers to infinity since they do not affect the global rate allocation. Then function $\Xi(\mathbf{Q}_p^j)$ transforms index set $(i,b)\in\mathbf{Q}_p^j$ to a linear index $k$ with relation $k=\sum_{\ell=1}^{i-1}L_\ell+b$.

To solve the above problem, we need to find the optimal association matrix $\boldsymbol{\eta}$ that maximizes the system's throughput given all users' decoding orders and the corresponding local rate allocations. Since all elements of $\boldsymbol{\eta}$ are 0 or 1, such problem is a 0-1 combinatorial optimization problem. Moreover, this problem cannot be solved efficiently using auction method~\cite{5089451}, since the rate for each layer in the summation is not fixed for each possible association matrix $\boldsymbol{\eta}$ because of Eq.~\eqref{eq:discard-layer-infty}. On the other hand, an exhaustive search is often prohibitive due to the large size of  matrix $\boldsymbol{\eta}$.
A feasible approach is to adopt the genetic algorithm~\cite{whitley1994genetic} to heuristically find a good solution. The details on the genetic algorithm are standard, and thus omitted in this paper.

\begin{algorithm}[htbp]
	\caption{Iterative rate update for $N_R$ users}
	\label{alg:iterative-rate-update-algo}
	\begin{algorithmic}[1]
		\State {Given channel-user association $\boldsymbol{\bar\eta}$, initial rate allocation $\bar{\mathbf{r}}$ and $N_R$ users at $\mathbf{x}_1,\ldots,\mathbf{x}_{N_R}$, set $\hat{\mathbf{r}}=\bar{\mathbf{r}}$;}
		\Repeat
		\For {$j=1$ to $N_R$}
		\State{$\mathcal{D}=\mathcal{K}_{\mathbf{x}_j}$, $\mathcal{G}=\emptyset$, $i=\underset{b}{\arg}\{ \eta_{jb}=1\}$, $p_j=0$;}
		\State{$\mathbf{r}^j[k]=+\infty, k=1,\ldots,\sum_{b=1}^{N_t}L_b$;}
		\State{$\mathbf{r}^j[\Xi((i,b))]=0, b=1,\ldots,L_i$;}
		\Repeat
		\State{ $\delta_\ell=\underset{\mathcal{V}\neq\emptyset,\mathcal{V}\subset\mathcal{D},|\mathcal{V}|\leq\tau}{\min}\Delta_{\mathbf{x}_j}(\mathcal{V},\mathcal{G},\hat{\mathbf{r}})$;}
		\State{$\mathcal{G}_\ell^{j}=\underset{\mathcal{V}\neq\emptyset,\mathcal{V}\subset\mathcal{D},|\mathcal{V}|\leq\tau}{\arg\min}\Delta_{\mathbf{x}_j }(\mathcal{V},\mathcal{G},\hat{\mathbf{r}})$;}
		\State{$\mathcal{D}\leftarrow\mathcal{D}\backslash\mathcal{G}_{\ell}^j$ and $\mathcal{G}\leftarrow\mathcal{G}\cup\mathcal{G}_{\ell}^j$;}
		\If{$\exists(i,b)\in\mathcal{G},b=1,\ldots,L_i$}
		\State{$\mathbf{r}^j[\Xi(\mathcal{G}_\ell^j)]=\delta_\ell$;}
		\State{$p_j\leftarrow p_j+1$;}
		\EndIf
		\State{$\ell=\ell+1$;}
		\Until {$D=\emptyset$}
		\State{Set $\mathbf{Q}_{m}^j\leftarrow\mathcal{G}_{\ell-m}^j,1\leq m\leq p_j$;}
		\State{ $\mathcal{Q}_{p_j+1}^j\leftarrow\cup_{m>p_j}\mathcal{G}_{\ell-m}^j$;}		
		\EndFor 
		\State{$\tilde{\mathbf{r}}[k]=\min_{j=1}^{N_R}\mathbf{r}^j[k], k=1,\ldots,\sum_{b=1}^{N_t}L_b$;}
		\State{$\hat{\mathbf{r}}[k]\leftarrow\hat{\mathbf{r}}[k]+\tilde{\mathbf{r}}[k], k=1,\ldots,\sum_{b=1}^{N_t}L_b$;}
		\Until{$\hat{\mathbf{r}}$ converges}
		\State{Output $\hat{\mathbf{r}}$ and $\{\mathcal{Q}^j\},j=1,\ldots,N_R$.}
	\end{algorithmic}
\end{algorithm}

Note that the above optimization problem provides a basic rate allocation and the associated decoding order. However, the sum rate gain provided by multi-layer coding and CPGD hasn't been fully exploited. An iterative rate update will be conducted at the transmitter side as detailed in Algorithm~\ref{alg:iterative-rate-update-algo}. The outputs of Algorithm~\ref{alg:iterative-rate-update-algo} are the enhanced rate allocation vector $\hat{\mathbf{r}}$ and the improved decoding orders $\mathcal{Q}^j$ for $j=1,\ldots,N_R$. The system can be configured through two phases. In the first phase, the transceivers employ the rate allocation and decoding orders obtained via solving the transmitter-user association problem, to establish the initial communication links. In the second phase, the transmitters distribute the updated decoding orders to the corresponding users, followed by updating the rate allocation for each transmission layer. The above two phases can enhance the system throughput promised by the proposed interference cancellation scheme.

\section{Numerical Results}
\label{sec:numerical-results}
\begin{figure}[htbp]
	\centering
	\subfigure[]{
	\includegraphics[width=0.4\linewidth]{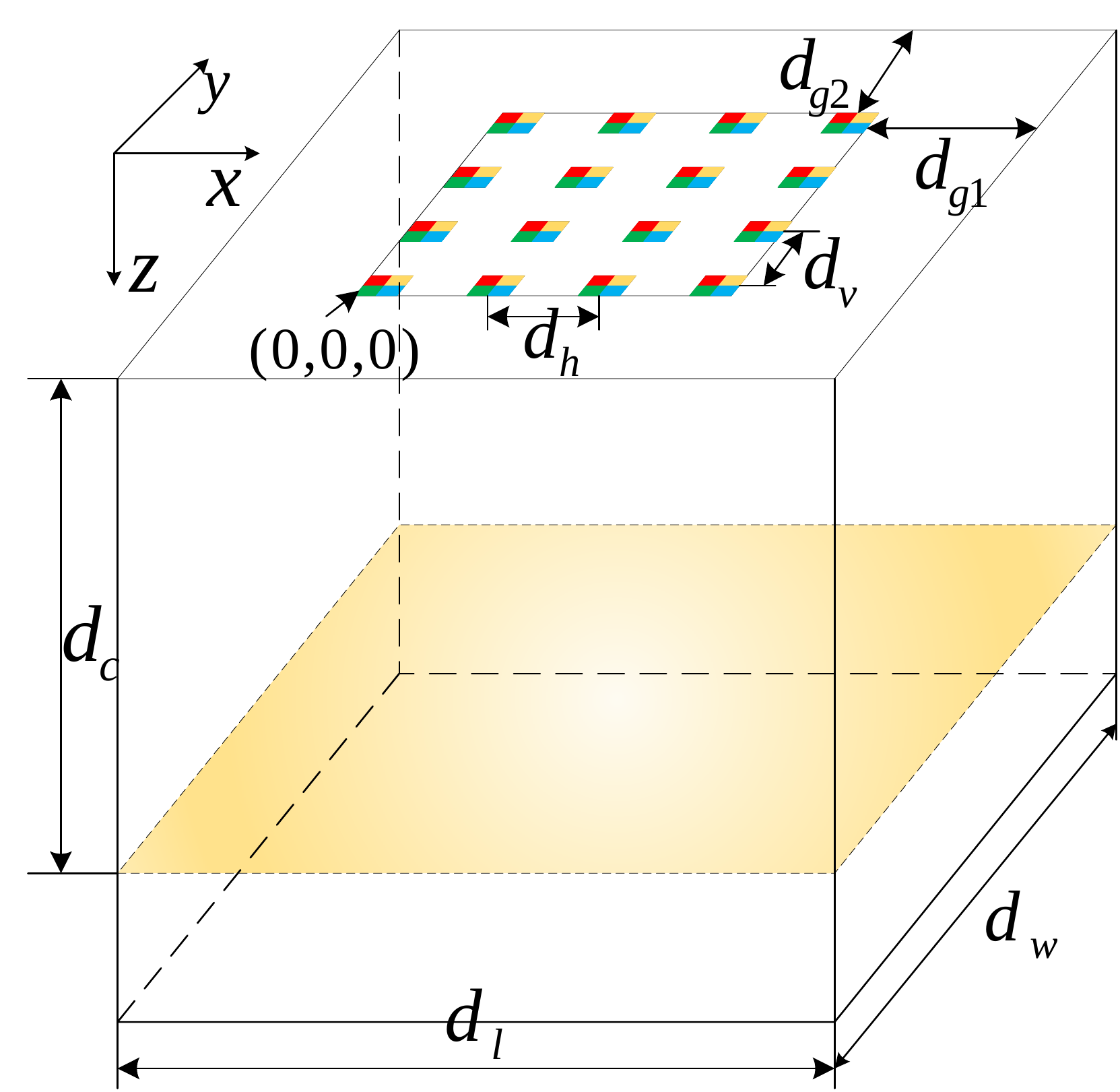}
	\label{fig:indoor-led-geometry}
	}
	\hfil
	\subfigure[]{
	\includegraphics[width=0.3\linewidth]{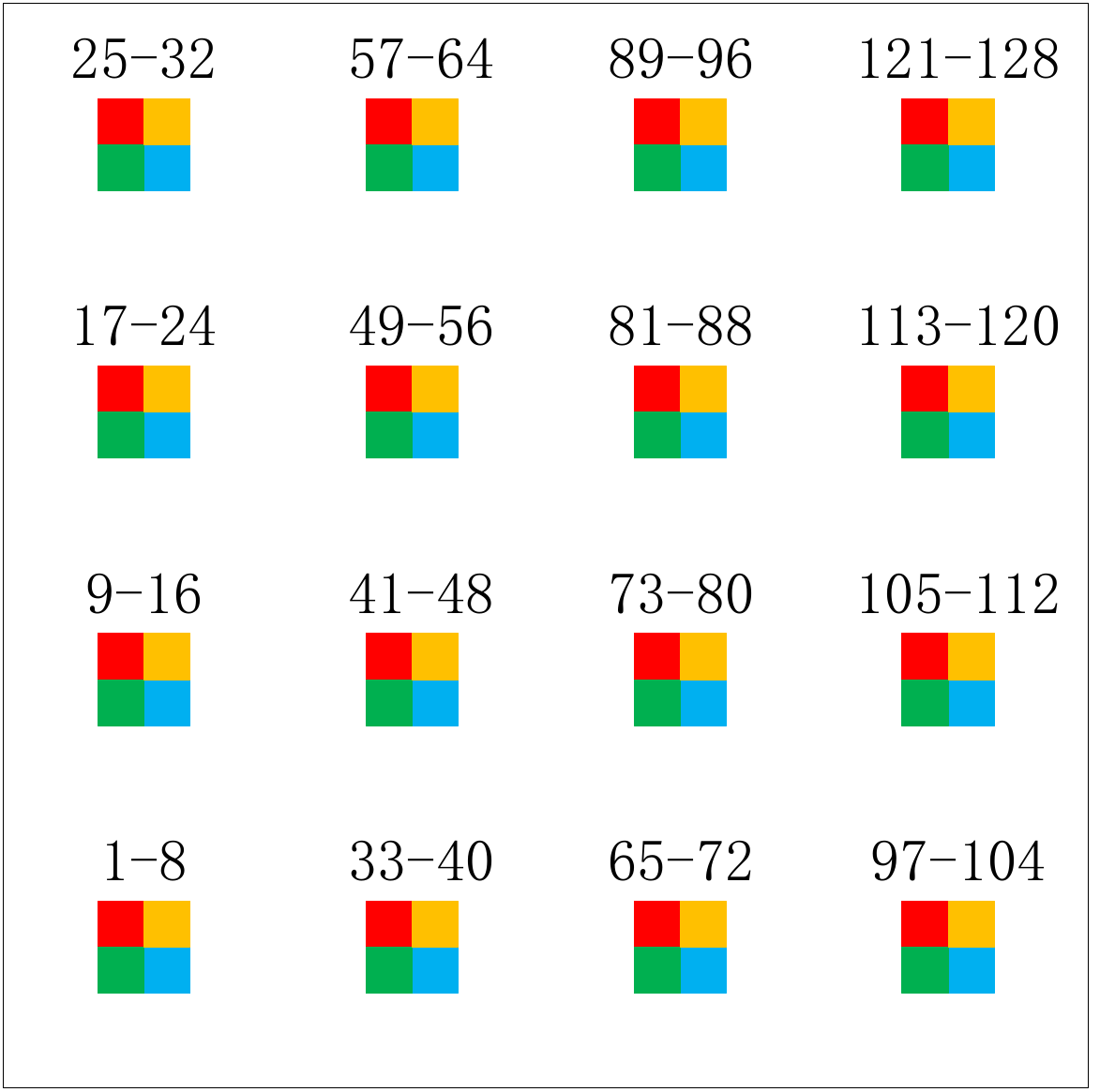}
	\label{fig:transmitter-layer-index}
	}
	\caption{(a) The indoor configuration for the multi-color multi-user VLC system under consideration (Unit: m); (b) LED indexes labeled sequentially for $4\times 4$ transmitters, each with four colors.}	
\end{figure}

Consider a practical indoor environment shown in Figure~\ref{fig:indoor-led-geometry}, with 16 transmitters with $4 \times 4$ squared layout with $d_h=d_v=0.2$m. At each transmitter, there are four LEDs with red, green, blue, and yellow colors whose spectra are shown in Figure~\ref{fig:four-color-spectrum}. The transmit array includes totally 64 equivalent transmitters, which is located at the center of the transmitter plane with margins $d_{g1}=d_{g2}=1$m. The receiver plane of interest consists of a plane with width $d_w$ and length $d_l$ both 2.6m, and is colored in yellow with vertical distance to the transmitter plane $d_c$ 2m. 

At the transmitter, the peak power of an LED is set to $A_i=1$ and the average power is set to $\varepsilon_i=0.5$, for all $i=1,\ldots,N_t$. The Lambertian order is $m_1=1$ with the transmitter semi-angle at half-power $\Phi_{1/2}=60^\circ$. At the receiver, a non-imaging detector is used with the active area $A_r=1\times10^{-4}\text{m}^2$ and the refractive index of optical concentrator $n=1.5$. Assume the receiver FOV $\psi_c=30^\circ$ (semi-angle). Let $h_{11}$ denote the channel gain from $(0,0,0)$ to $(0,2,0)$ through the optical filter with passband [380nm, 480nm]. The noise variance $\sigma^2$ is set such that the receiver-side SNR $A_ih_{11}/\sigma=15$dB, which is typical for indoor VLC with strong light intensity. The noise variance is assumed to be the same for all positions on the receiver plane.

Multi-layer encoding and CPGD are employed at the transmitter and the receiver, respectively. For each transmitter, two-layer encoding is adopted in the simulation to achieve the tradeoff between multi-layer coding gain and computational complexity. The layer indexes are labeled sequentially in Figure~\ref{fig:transmitter-layer-index} for illustration. At each position, the layer index increases from  the LED with the smallest wavelength to that with the largest wavelength. For each layer, the input signal obeys a TG distribution defined in Eq.~\eqref{eq:truncated-gaussian} with $\mu_{ik}$ and $\nu_{ik}$ to be determined. We choose $\mu_{ik}=3\nu_{ik}$ as suggested in~\cite{ChaabanVLCbounds}, and combine Eq.~\eqref{eq:opt-constraints} and Eq.~\eqref{eq:tg-mean} to compute $\hat{\mu}_{ik}$ and $\hat{\nu}_{ik}$ for the TG distribution. For each receiver, a CPGD with group size one is employed, since it has been reported that the group size of one in CPGD has a low complexity to achieve the most of the performance gain, especially in the scenario with channel coding~\cite{gong2011interference}.


To obtain the decoding map, Algorithm~\ref{alg:dec-order-pos-algo} is adopted to compute the decoding order and local rate allocation at each position uniformly sampled with the sampling gap $d_s=0.1\text{m}$ on the receiver plane, which are then stored.  Note that any of the four colors can be adopted by a user. 

\begin{figure}[htbp]
	\centering
	\subfigure[Decoding order]
	{
		\includegraphics[width=0.46\linewidth]{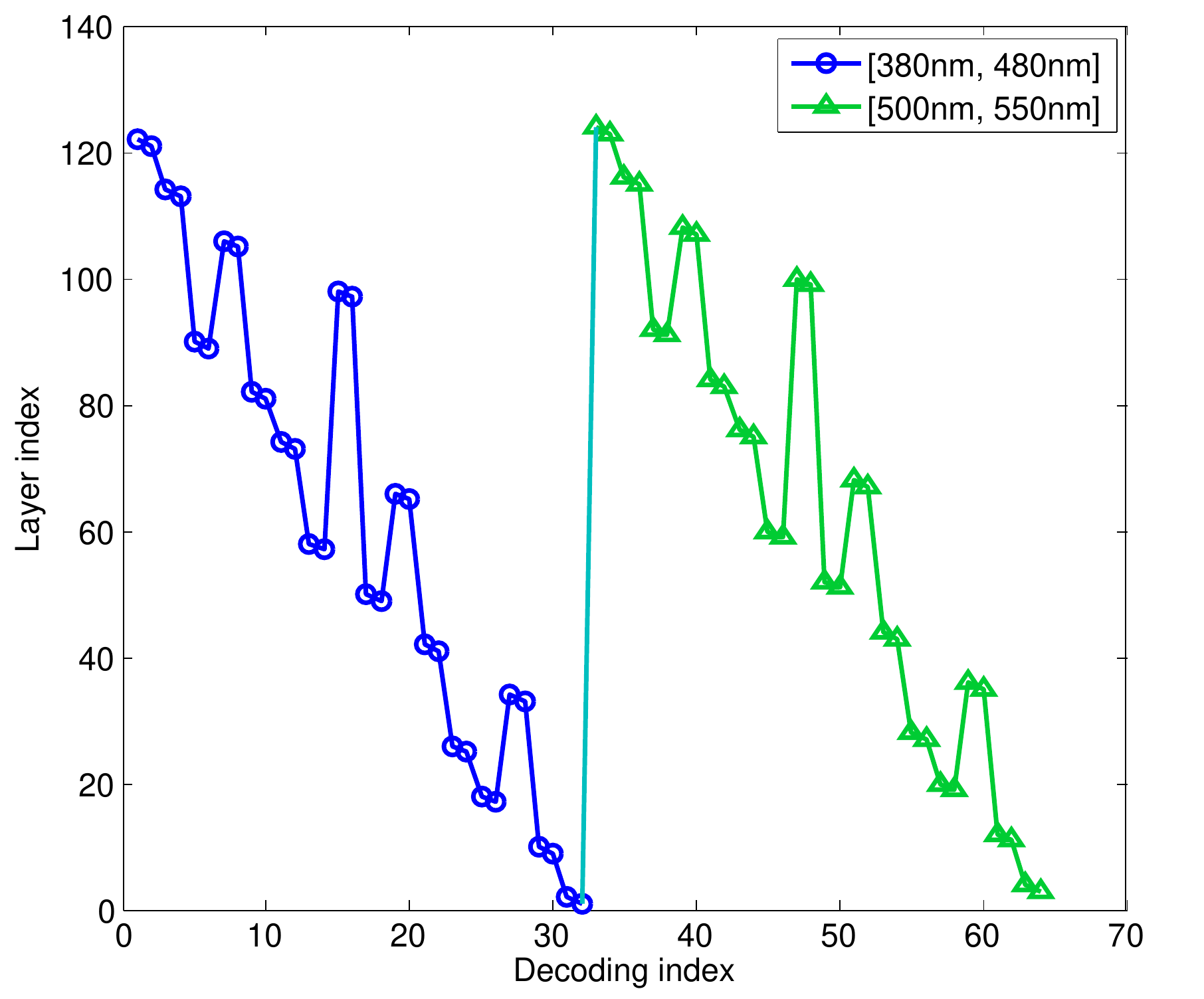}\label{fig:decoding-sequence-illustration}
	}	
	\hfil
	\subfigure[Maximum achievable rates for all layers]
	{
		\includegraphics[width=0.46\linewidth]{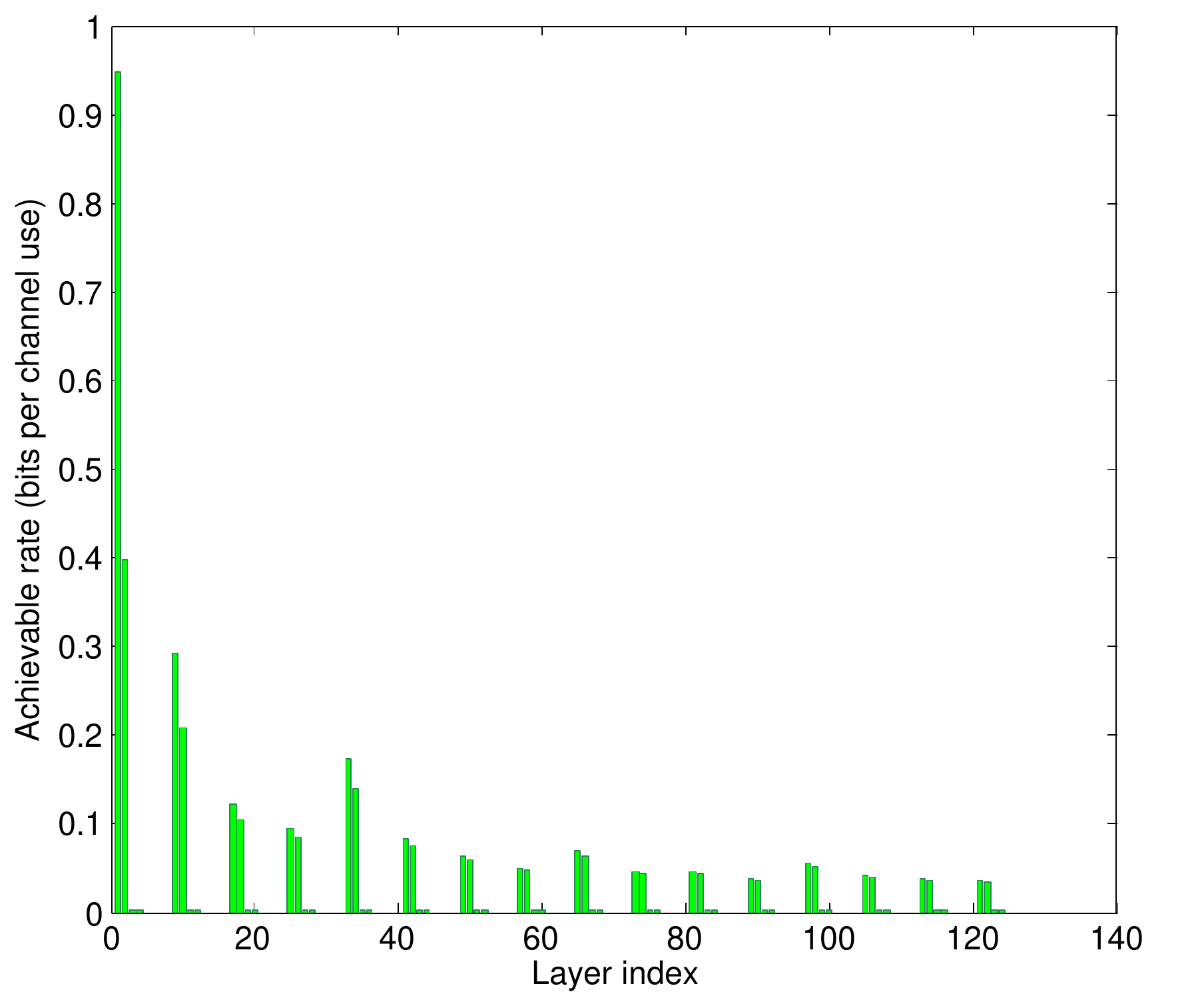}\label{fig:rate-allocation-illustration}
	}
	\caption
	{
		The decoding order and local rate allocation at position $(0.8,0.6,2)$ of the decoding map obtained from the signal passing the optical filter with passband [380nm, 480nm].		
	}
	\label{fig:decoding-sequence-and-rate-allocation}
\end{figure}

Figure~\ref{fig:decoding-sequence-and-rate-allocation} gives the result at position $(0.8,0.6,2)$ of the decoding map constructed from the signal passing the optical filter with passband [380nm, 480nm]. Figure~\ref{fig:decoding-sequence-illustration} shows the signal layer decoded at each decoding stage. From Figure~\ref{fig:decoding-sequence-illustration} we see that layers sent from the same color are decoded consecutively and layers from color with wavelength [500nm, 550nm] will not be decoded until all the layers from other colors are recovered successfully. The reason is the large difference in signal strength induced by optical filters. The layers from color [560nm, 600nm] and color [600nm, 680nm] are not presented since the intensity of signal passing through the optical filter with passband [380nm, 480nm] is negligible. Figure~\ref{fig:rate-allocation-illustration} presents the maximum achievable rates for all layers when the decoding order is given in Figure~\ref{fig:decoding-sequence-illustration}. This gives the local rate allocation for the layers at position $(0.8,0.6,2)$.

\begin{figure}[htbp]
	\centering
	\includegraphics[width=0.5\linewidth]{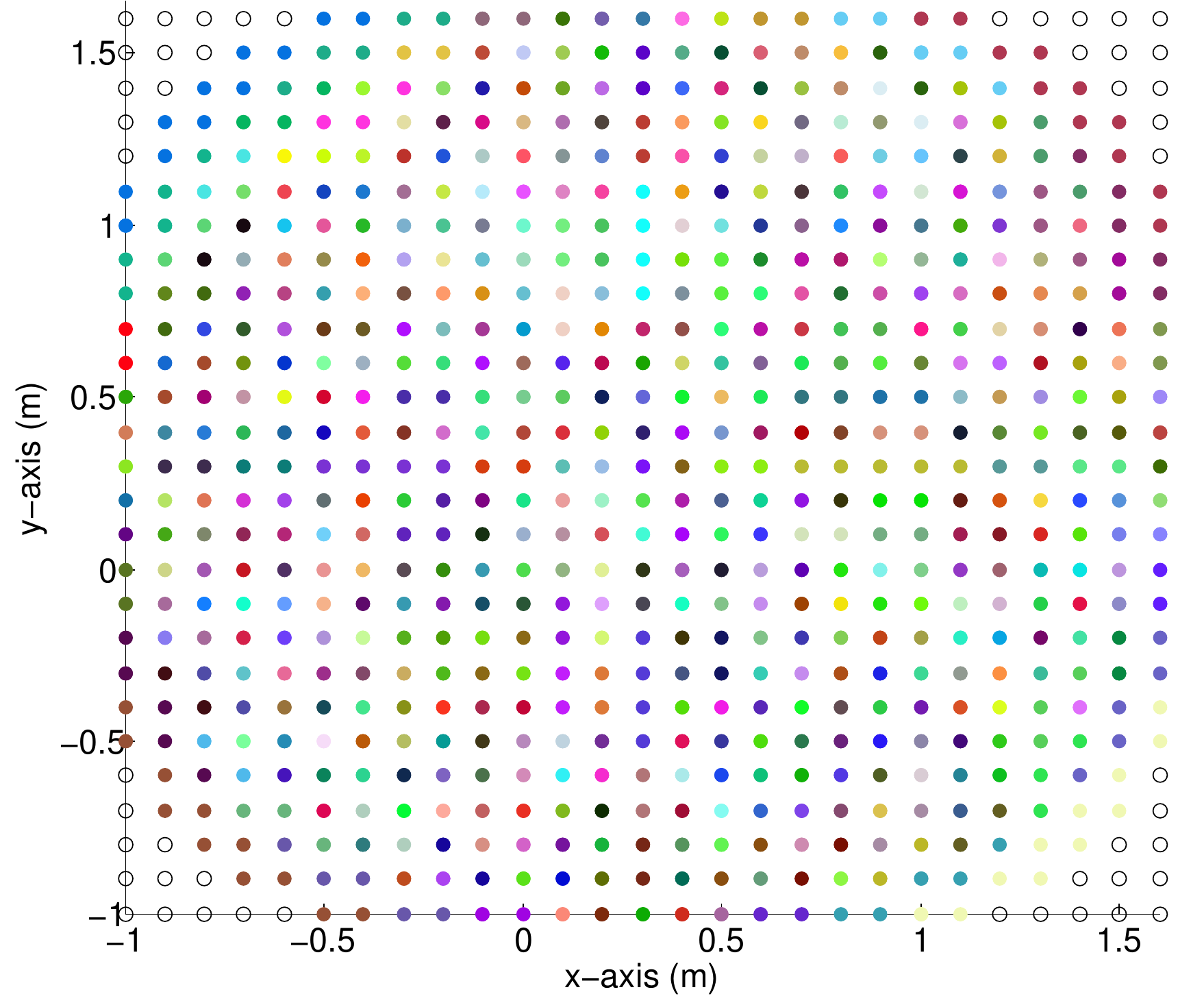}
	\caption{Clustering of the decoding map with $4\times4$ transmitters, each with four colors, obtained at color band [380nm, 480nm].}
	\label{fig:clustering-16tx}
\end{figure}

For each decoding map of a receiver plane, we employ Algorithm~\ref{alg:size-reduced-algo} to reduce its size. In our simulation, $\tau_{\text{diff}}$ was set to $1\times10^{-5}$ and $\tau_{\text{loss}}$ was set to 0.1. Figure~\ref{fig:clustering-16tx} shows the clustering result of the decoding map of color [380nm, 480nm] under the indoor setting shown in Figure~\ref{fig:indoor-led-geometry}. The black empty circles represent the positions that will yield decoding outages because of the weak signal reception. The filled circles with the same color are in the same cluster denoting that they share the same decoding order. In this map, a total of $N_s=681$ points are classified into $N_k=511$ clusters, and the compression ratio defined as $(N_s-N_k)/N_s$ is 25.0\%. The maximum average loss (or the maximum average normalized distance) is close to zero, meaning that sharing the same decoding order will not penalize the system's sum rate.

\begin{figure}[htbp]
	\centering
	\includegraphics[width=0.5\linewidth]{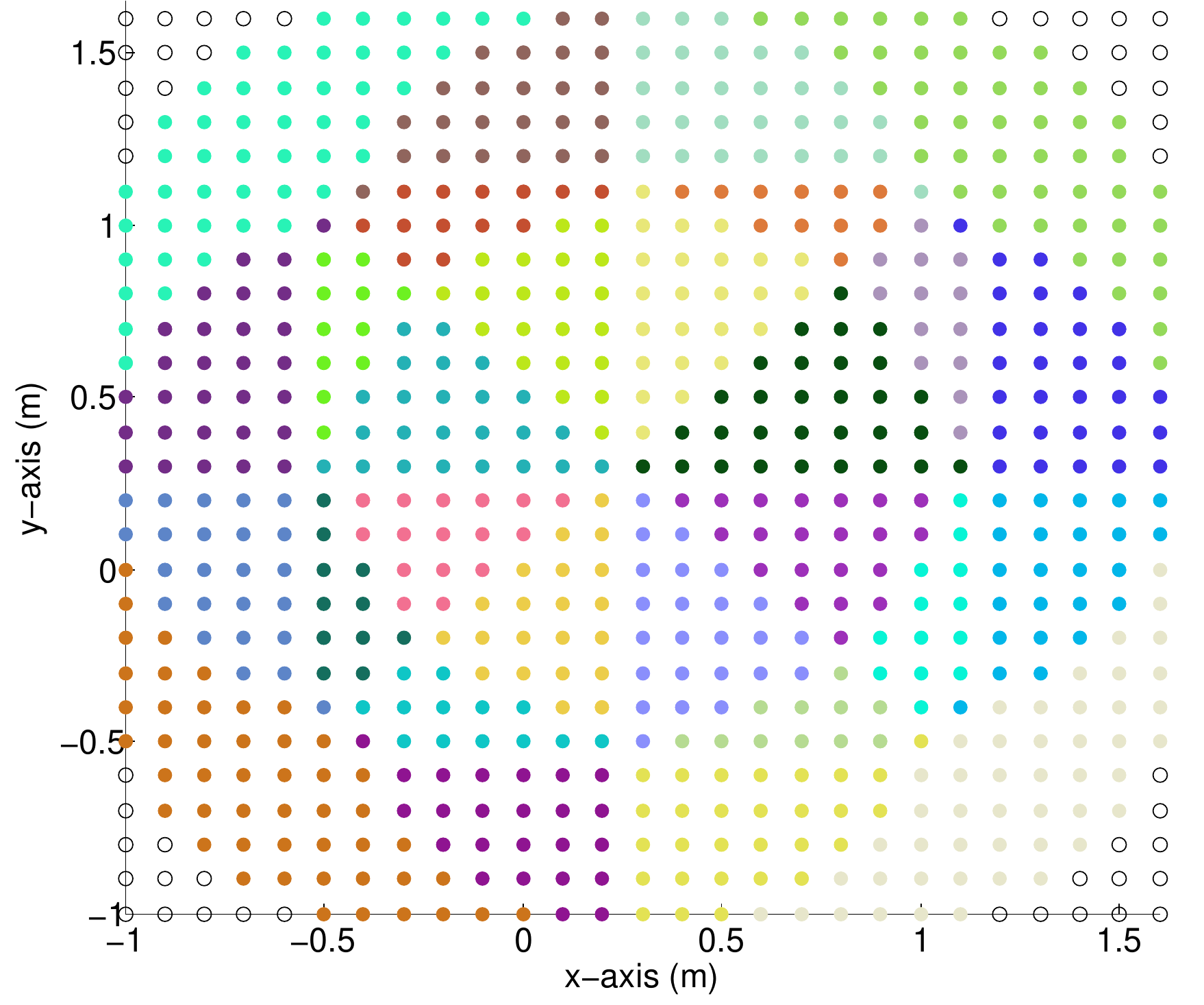}
	\caption{Clustering of the decoding map with $2\times2$ transmitters, each with four colors, obtained through an optical filter with passband [380nm, 480nm].}
	\label{fig:clustering-4tx}
\end{figure}

\begin{figure}[htbp]
	\centering
	\includegraphics[width=0.5\linewidth]{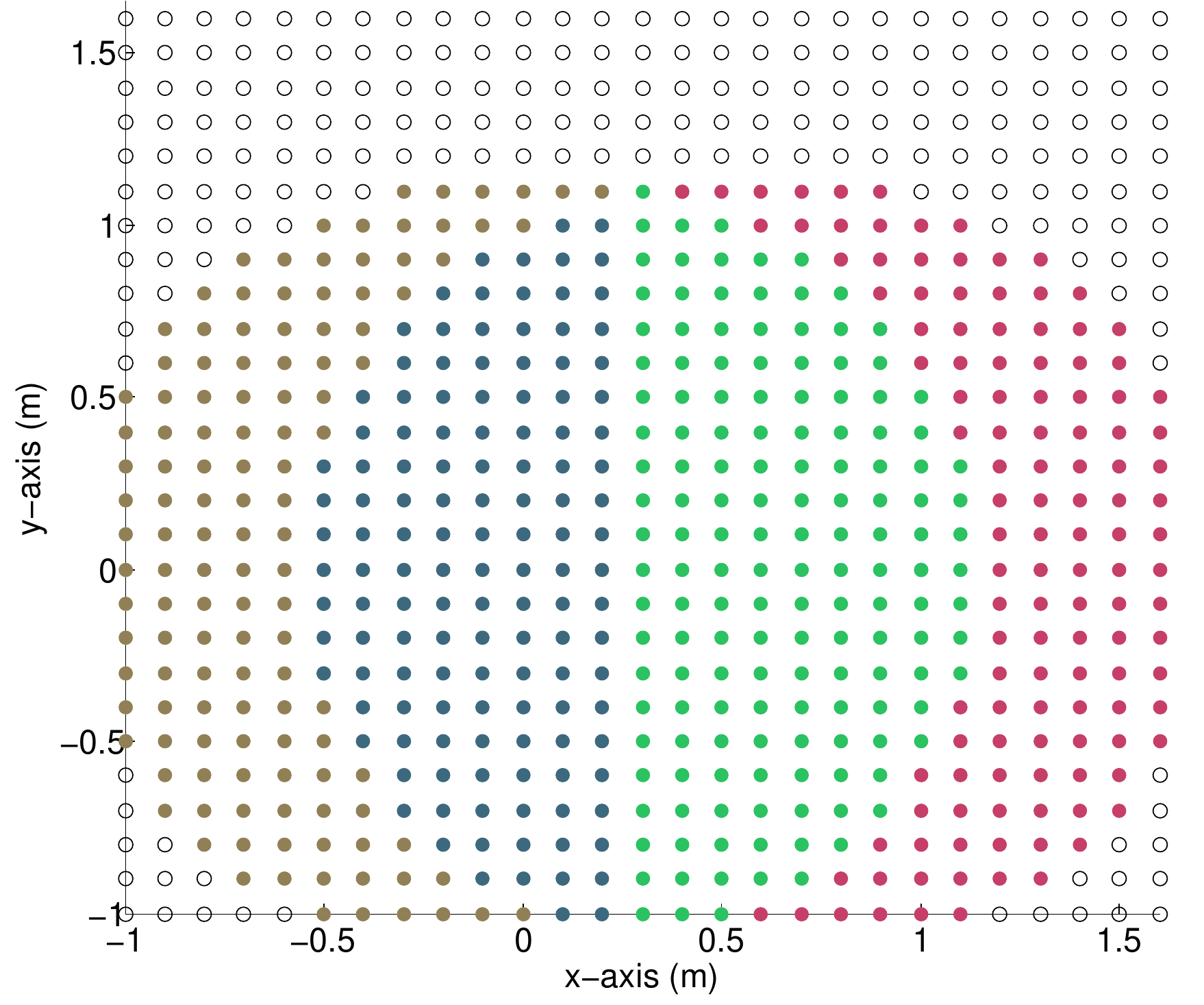}
	\caption{Clustering of the decoding map with $1\times2$ transmitters, each with four colors, obtained through an optical filter with passband [380nm, 480nm].}
	\label{fig:clustering-2tx}
\end{figure}

We show that the density of the transmitters affects the effectiveness of Algorithm~\ref{alg:size-reduced-algo}. The compression ratio will increase if we reduce the number of transmitters within the same area. Figure~\ref{fig:clustering-4tx} gives the clustering result of the decoding map with only four transmitters at four corners located at $(0,0,0)$, $(0,0.6,0)$, $(0.6,0,0)$, $(0.6,0.6,0)$. The decoding map has been clustered into 28 categories with the maximum average loss close to zero and the compression ratio being 95.9\%, which demonstrates the effectiveness of the size reduction algorithm. Figure~\ref{fig:clustering-2tx} gives the clustering result when transmitters are located at $(0,0,0)$ and $(0,0.6,0)$. The decoding map has been clustered into 4 categories with maximum average loss close to zero and the compression ratio being 99.2\%. 


Note that Algorithm~\ref{alg:size-reduced-algo} uses the distance defined in~\eqref{eq:normalized-distance} to characterize the similarity between two decoding orders, to perform clustering and reduce the size of a decoding map. But it ignores the possible symmetry of channel gains, which can establish an explicit relation between two decoding orders at two symmetrical positions. As demonstrated in Section~\ref{sec:symmetry-constraint-size-reduction}, the transmitters arranged in a square shape will have symmetrical positions having the same channel gain set on the receiver plane, which shows an explicit relationship between the corresponding two decoding orders. The compression ratio of Figure~\ref{fig:clustering-16tx}  with a higher density of transmitters can be further increased to $7/8$ or 87.5\% if we only compute the upper right triangle region of the decoding map and obtain the rest  based on that region. The number of decoding orders needed to be stored in Figure~\ref{fig:clustering-4tx} can be reduced from 28 to 7 yielding an increased compression ratio of 99.0\%.

We further pursue the optimal transmitter-user association and rate allocation to maximize the system throughput. In the following, we consider users using color band [380nm, 480nm] as an example. The decoding map we use in the following simulations is constructed from the symmetric properties defined in Section~\ref{sec:symmetry-constraint-size-reduction}, i.e. we only compute the upper right 1/8 area of the whole decoding map, and construct the remain 7/8 area based on the  1/8 results.

\begin{figure}[htbp]
	\centering
	\includegraphics[width=0.5\linewidth]{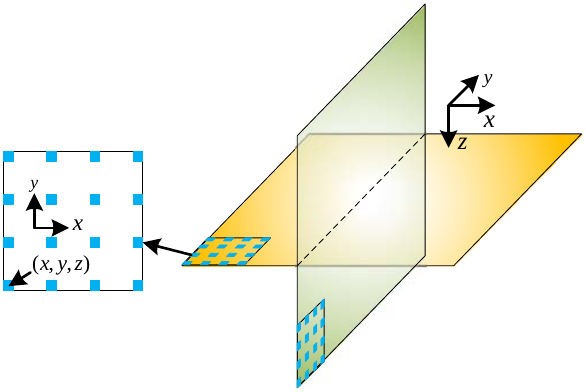}
	\caption{Illustration of 16 users operating on the same color spectrum [380nm, 480nm] moving on the $x-y$ and $y-z$ receiver planes.}
	\label{fig:users-moving-plane}
\end{figure}

Figure~\ref{fig:users-moving-plane} gives positions of 16 users uniformly over the receiver plane either parallel or perpendicular to the transmitter plane in a squared grid layout with the distance between adjacent users being 0.2m. On the $x-y$ plane, the position of the bottom left user $(x,y,z)$ with $z=2$ represents all the 16 users' positions, which are constrained in the region $-1\leq x\leq 1.6$ and $-1\leq y\leq 1.6$. On the $y-z$ plane, the position of the bottom left user $(x,y,z)$ with $x=0.3$ represents all the 16 users' positions. Users are confined in region $-1\leq x\leq 1.6$ and $1\leq z\leq 3$.  Then on each plane, we move all the users via step 0.1m, and solve 
the maximization problem of Eq.~\eqref{eq:chan-assoc-rate-alloc-problem} subject to the constraints from Eq.~\eqref{eq:unique-rx-constraint} to Eq.~\eqref{eq:initial-constraint} by genetic algorithm. 

\begin{figure}[htbp]
	\centering
	\includegraphics[width=0.5\linewidth]{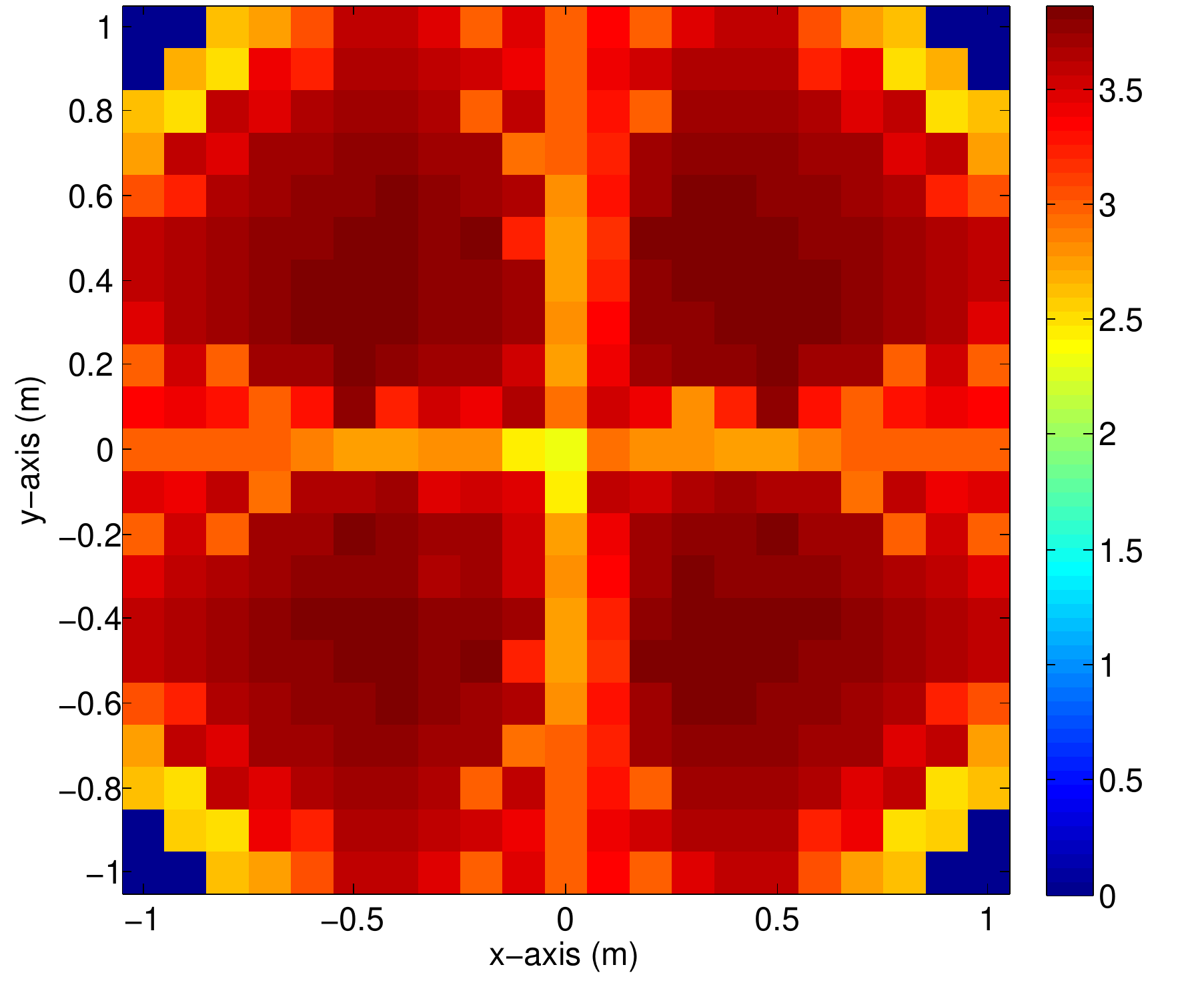}
	\caption{Sum rate of 16 users operating at color [380nm, 480nm] moving over the $x-y$ receiver plane with transmitter-user association.}
	\label{fig:sum-rate-plane-xy}
\end{figure}
\begin{figure}[htbp]
	\centering
	\includegraphics[width=0.5\linewidth]{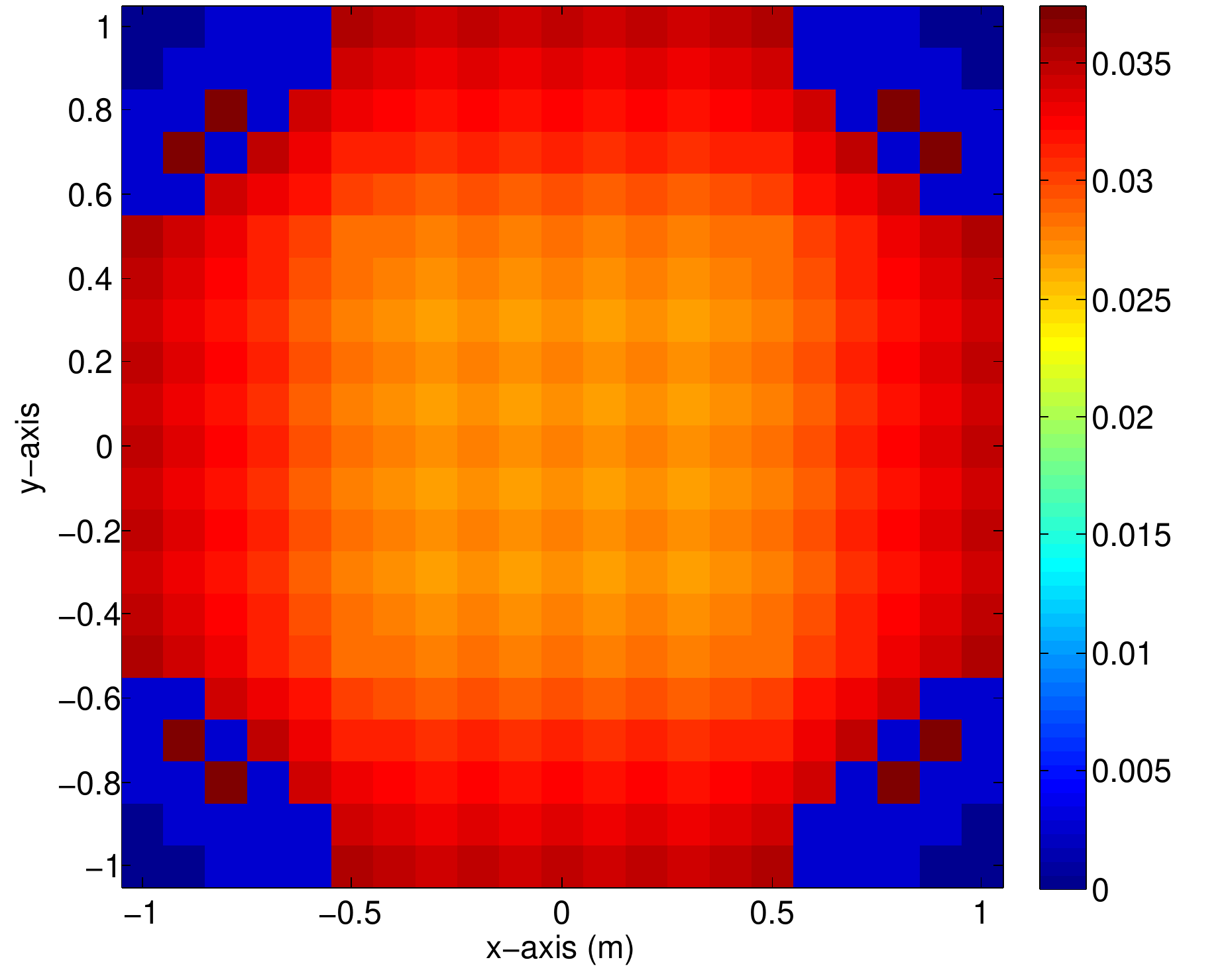}
	\caption{Minimum rate of 16 users operating at color [380nm, 480nm] moving over the $x-y$ receiver plane with transmitter-user association.}
	\label{fig:min-rate-plane-xy}
\end{figure}

Figure~\ref{fig:sum-rate-plane-xy} gives the sum rate on the $x-y$ receiver plane parallel to the transmitter plane when the transmitter-user association is taken into account and iterative rate update is conducted. We observe lower rates at $x = 0$ and $y = 0$
than those of their adjacent positions. This can be justified as follows. Assuming the associated transmitter for user $u_1$ is $i_1$, then for user $u_2$, which is symmetrical to $u_1$, the associated transmitter is $i_2=\mathbf{A}_p(\pi^{-1}_{\mathbf{A}}(i_1))$. Then at $u_1$, generally we have $R_1>R_2$, where $R_1=\mathbf{r}^{u_1}[\Xi((i_1,k))]$ and $R_2=\mathbf{r}^{u_1}[\Xi((i_2,k))]$; at $u_2$, by symmetry, we have $R_1'=R_2$, $R_2'=R_1$ with $R_1'=\mathbf{r}^{u_2}[\Xi((i_1,k))]$ and $R_2'=\mathbf{r}^{u_2}[\Xi((i_2,k))]$. By taking the element-wise minimum operation to $\mathbf{r}^{u_1}$ and $\mathbf{r}^{u_2}$, the rates of layers $(i_1,k)$ and $(i_2,k)$ are both constrained to the lower rate $R_2$. Thus strong symmetric interference exhibited by user locations may constrain the throughput of the system. Moreover, it is observed that based on the allocation results, each user is allocated to the colors matching the corresponding wavelengths.

On the other hand, if the majority of users are at asymmetrical positions, this rate constraint will be lessened since some other user may suffer weaker interference for the same transmitter layer. So at the center of the receiver plane, the sum rate is the lower and it increases as users move off the center, which finally decreases due to the attenuated signal strength.
\begin{figure}[htbp]
	\centering
	\includegraphics[width=0.5\linewidth]{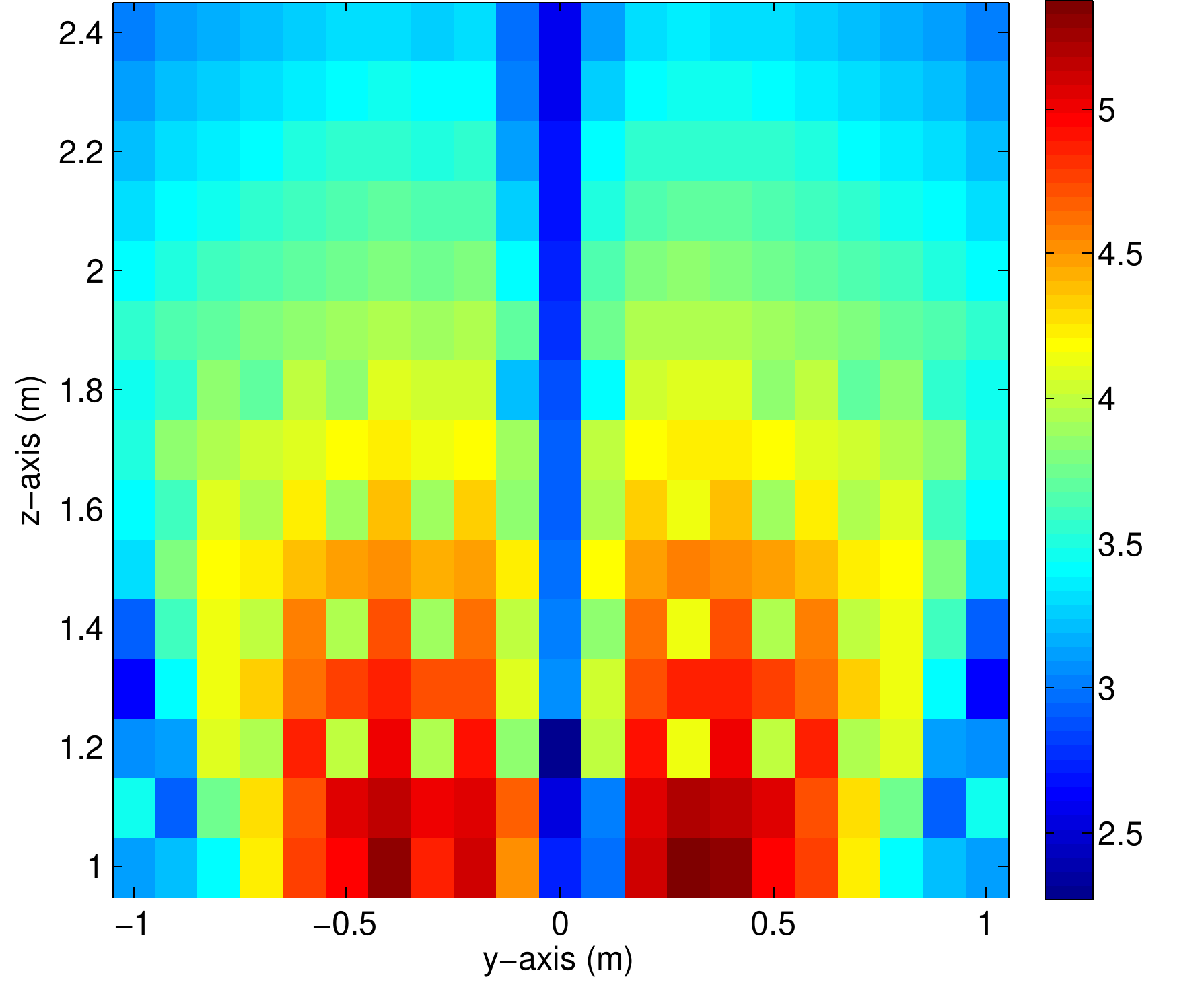}
	\caption{Sum rate of 16 users operating at color [380nm, 480nm] moving over the $y-z$ receiver plane with transmitter-user association.}
	\label{fig:sum-rate-plane-yz}
\end{figure}
\begin{figure}[htbp]
	\centering
	\includegraphics[width=0.5\linewidth]{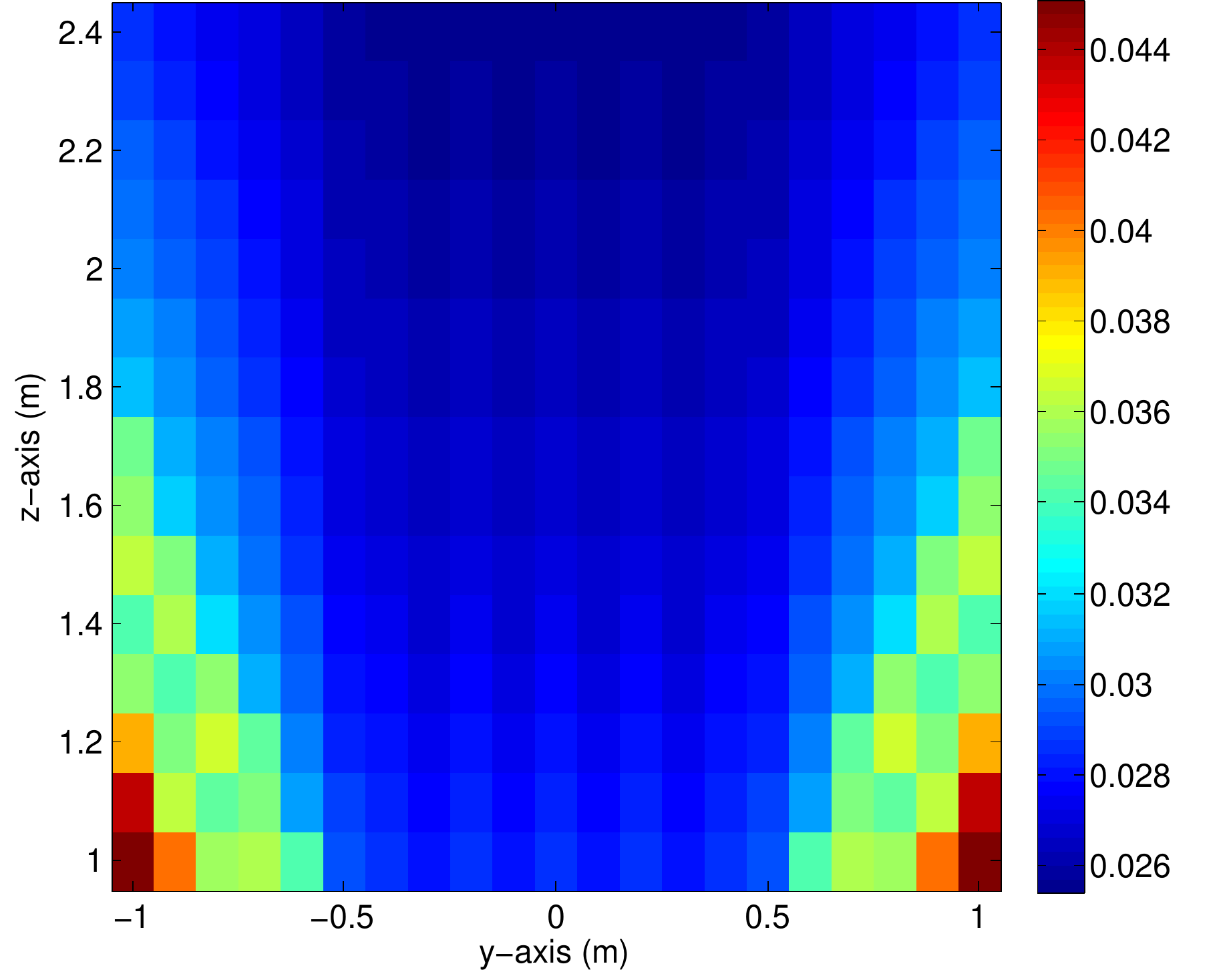}
	\caption{Minimum rate of 16 users operating at color [380nm, 480nm] moving over the $y-z$ receiver plane with transmitter-user association.}
	\label{fig:min-rate-plane-yz}
\end{figure}
Figure~\ref{fig:min-rate-plane-xy} gives the minimum rate distribution over the $x-y$ plane. Similarly, it is observed that the minimum rate first increases and then decreases as user matrix moves off the center, due to the interference structure and attenuated link gain, respectively. 

Figure~\ref{fig:sum-rate-plane-yz} gives the sum rate distribution on the $y-z$ receiver plane perpendicular to the transmitter plane.  Similarly, we observe lower rates at $y=0$ compared with those of adjacent positions. The number of symmetrical users is highest at $y=0$, which constrains the throughput. 
The sum rate increases as the users move to left or right positions and decreases as the signal strength decays. Furthermore,
we observe similar rate variation in Figure~\ref{fig:min-rate-plane-yz} on the minimum rate distribution over the $y-z$ plane.


\section{Conclusion}
\label{sec:conclusion}
In this paper, we have employed multi-layer coding and CPGD for interference cancellation in VLC, where signals are transmitted via multiple LEDs with different colors. We model the multi-color multi-user interference channel with Gaussian-like spectra and ideal optical filters.  A max-min fairness optimization problem is formulated, based on the achievable rate of a VLC-MAC obtained from truncated Gaussian distribution. The problem is solved by greedy algorithm with no preset transmitter-user associations. A decoding map is constructed to reduce the online computational complexity. A classification-based algorithm is proposed to reduce the map size and facilitate storage and searching, where the compression ratio can be further increased based on the symmetrical geometry of the transmitters. Finally, the transmitter-user association problem is formulated as a 0-1 optimization problem solved by genetic algorithm, followed by an iterative update of the rate allocations and decoding orders. The sum and minimum achievable rates are evaluated by numerical results.

\ifCLASSOPTIONcaptionsoff
  \newpage
\fi



\bibliographystyle{IEEEtranTCOM}
\bibliography{vlc_cap}
%
%
%
%

%




\end{document}